\documentclass[11pt,a4paper]{article}
\usepackage{jheppub}
\usepackage{graphicx}
\usepackage{lipsum}
\usepackage{slashed}
\usepackage{mathtools}
\usepackage{bm}
\usepackage[all]{hypcap}
\usepackage{tikz}
\usepackage{tikz-feynman}
\usepackage{bbm}
\usepackage{multirow}
\usepackage{xcolor}
\usepackage{color, colortbl}
\usepackage{subcaption}
\usepackage{enumitem}
\usepackage{cancel}
\usepackage{soul}
\usepackage[font=footnotesize,labelfont=bf]{caption}

\allowdisplaybreaks

\def\Z{\mathbb{Z}}
\def\R{\mathbb{R}}

\def\ZP{Z^\prime}
\def\WP{W^\prime}

\def\g{G}
\def\h{H}
\def\so{{\rm SO}}
\def\sp{{\rm Sp}}
\def\sm{H_{\rm SM}}
\def\su{{\rm SU}}
\def\u{{\rm U}(1)}
\def\su{{\rm SU}}
\def\u{{\rm U}(1)}
\newcommand{\UU}{\mathrm{U}}
\def\heavy{{[3]}}
\def\light{{[12]}}
\def\productn{\!\times\!}
\def\product{\times}

\newcommand{\cL}{{\cal L}}

\newcommand{\gp}{g'_{_{W}}}

\newcommand{\rep}[1]{\mathbf{#1}}
\newcommand{\repbar}[1]{\overline{\mathbf{#1}}}

\usepackage{xspace}

\newcommand{\be}{\begin{equation}}
\newcommand{\ee}{\end{equation}}
\newcommand{\bea}{\begin{eqnarray}}
\newcommand{\eea}{\end{eqnarray}}
\newcommand{\beq}{\begin{equation}}
\newcommand{\eeq}{\end{equation}}

\begin{document}

\title{Non-universal gauge interactions addressing 
the inescapable link between Higgs and Flavour}

\author{Joe Davighi and Gino Isidori}

\affiliation{Physik-Institut, Universit\"at Z\"urich, CH 8057 Z\"urich, Switzerland}
\emailAdd{joe.davighi@physik.uzh.ch}
\emailAdd{isidori@physik.uzh.ch}

\abstract{We systematically explore ultraviolet complete models where flavour hierarchies emerge, via approximate accidental 
symmetries, from an underlying flavour non-universal gauge structure. In order to avoid large quantum corrections to the Higgs mass, 
the first layer of non-universality, separating the third generation from the light ones, should appear at the TeV scale. 
A handful of models survive the combined criteria of naturalness in the Higgs sector, having a semi-simple embedding in the UV, and 
compatibility with experiments. They all feature quark-lepton unification in the third family and a non-universal electroweak sector. 
We study in more detail the interesting option of having colour and hypercharge non-universal at the TeV scale, while 
 $\mathrm{SU}(2)_L$ remains universal up to high scales: this gauge structure turns to be very efficient in secluding the Higgs from large 
quantum corrections and predicting  flavour mixing consistent with data.  In all cases, these models imply a rich TeV-scale phenomenology 
within the reach of near-future direct and indirect experimental searches.}

\maketitle

\section{Introduction}

The stability of the Higgs mass under quantum corrections and the non-trivial structure of the Higgs--fermion Yukawa couplings
(often referred to as the flavour puzzle)
are two central problems in particle physics. Neither signals an internal inconsistency of the Standard Model (SM), but rather a peculiar tuning of its parameters. Barring the possibility that this tuning is accidental, or that it finds an explanation outside the realm of Quantum Field Theory (QFT), inevitably leads us to the conclusion that the SM has a non-trivial ultraviolet (UV) completion as a QFT.
From this perspective, which is the one we adopt in this paper, these two problems are naturally linked.

Explaining the flavour structure of the SM in terms of some underlying dynamics requires extra degrees 
of freedom in the UV that couple to the Higgs field. These degrees of freedom unavoidably contribute to 
destabilise the Higgs mass. We stress that this is not a generic statement about the Higgs hierarchy problem
({\em i.e.}~the generic quadratic sensitivity of the Higgs mass terms to UV physics), but a precise (calculable) 
effect that we can quantify in QFT~\cite{Farina:2013mla}. Given that some of the Yukawa couplings are of $O(1)$, 
if the UV physics addressing the flavour problem appears at high scales, this destabilisation is likely 
to produce a serious fine tuning among physical scales.  An analogous conclusion holds 
if we aim to explain the seemingly {\em ad hoc} pattern of quantised SM hypercharges by embedding the SM gauge symmetry into a larger group, as in the classical  models of grand unification~\cite{Georgi:1974sy,Pati:1974yy}.

A large fraction of the model-building attempts proposed so far have tried to disentangle these two problems
using a general scale separation.  The basic idea was to stabilise the Higgs sector 
just above the electroweak scale, via some sort of flavour-universal new physics (such as supersymmetry or
Higgs-compositeness), while postponing the origin of flavour dynamics, and possibly unification, to higher 
scales.\footnote{This scale separation is possible because the flavour and hypercharge-quantization problems 
arise by marginal operators, not associated to a well-defined energy scale.}
This approach, which was a very natural option in the pre-LHC era, is not so well motivated nowadays:
the absence of direct signals of new physics has pushed well above 1~TeV the bounds on new degrees of freedom with 
$O(1)$  flavour-universal couplings to the SM fields, worsening the tuning on the Higgs mass. 
However, importantly, these stringent bounds are not derived from direct couplings of the new physics to the 
Higgs or the top quark, but rather by its couplings to the light SM fermions, which play a minor role in the stability of the Higgs mass. A similar conclusion applies for the indirect 
constraints on new physics couplings that come from precision measurements: it is the new physics couplings to the light SM fermions
that are strongly constrained, not those to the third generation. 

This phenomenological observation prompts us to consider a different approach to address the Higgs and flavour problems:
rather than separating them, we try to address them at the same time via flavour non-universal gauge interactions at the TeV scale.  
The idea of addressing the origin of the  flavour hierarchies via fundamentally flavour non-universal gauge interactions, that couple most strongly to the third family, has attracted a lot of attention in the last few years~\cite{Bordone:2017bld,Greljo:2018tuh,Fuentes-Martin:2020pww,Fuentes-Martin:2020bnh,Fuentes-Martin:2022xnb,FernandezNavarro:2022gst,Davighi:2022bqf}.  
All these recent attempts have been phenomenologically motivated by  the 
so-called $B$-physics anomalies, but their interest goes well beyond them. One common aspect shared by these models is the {\em a posteriori}
justification of an approximate $\UU(2)^5$ flavour symmetry acting on the light fermion families, which is known to provide a good description of the SM spectrum plus an efficient suppression of flavour-violating effects for low-scale 
new physics~\cite{Barbieri:2011ci,Isidori:2012ts}. By invoking flavour non-universal gauge interactions,  the $\UU(2)^5$ symmetry emerges as an accidental symmetry of the gauge sector.  
This approach also connects to the older, and more general, 
idea of a UV completion of the  SM based on a (flavoured) multi-scale construction~\cite{Dvali:2000ha,Panico:2016ull,Allwicher:2020esa,Barbieri:2021wrc}.
In this context the naturalness of the Higgs mass can be preserved, or, more accurately, the tuning is minimized, if the first layer of new physics enters near the TeV scale, with subsequent layers separated by a few orders of magnitude~\cite{Allwicher:2020esa}.
This of course does not rule out the possibility that supersymmetry or compositeness 
play a role in screening the scalar sector from dynamics at even heavier scales; however, if utilised, these general stabilisation mechanisms 
could manifest at higher scales given the low-scale stabilisation provided by the lighter flavour non-universal layer of new-physics.

In this work, we undertake a general study of flavour non-universal gauge dynamics at the TeV scale, without relying on any hint of deviations from the SM in current data. We look for models that fully explain the SM flavour structure, via an approximate $\UU(2)^5$ flavour symmetry  emerging accidentally  at the TeV scale, plus  
$\UU(2)^5$-breaking effects that appear as a remnant of further non-universal dynamics at higher scales that is screened from the Higgs. 
All the models are ultimately completed by semi-simple gauge theories in the UV, thereby shedding light also on the hypercharge quantisation problem via unification, albeit in an intrinsically flavour non-universal manner. By estimating the corrections to the Higgs mass, we use naturalness as a guide to steer our attention to the most natural explanations of flavour, consistent with complementary constraints from flavour observables (which render theories with flavour universal quark-lepton unification to be unnatural, for example) and from the non-observation of proton decay (which rules out models with, say, $\su(5)$-based unification schemes remotely close to the TeV scale).

As we shall show, there is a handful of models satisfying all these conditions, some of which have never been considered in previous literature. 
One of the most interesting aspects is that all viable models share common features, which allow us to deduce some general phenomenological
conclusions that follow from our main hypotheses. These include the expectation of heavy-flavoured gauge bosons at the TeV scale, coupled mainly to the third family, necessarily including the $U_1$ vector leptoquark. In addition, at least one family of vector-like fermions is expected near the TeV scale. 
The most successful model satisfying all our naturalness requirements is a model where  $\su(2)_L$ remains flavour universal up to high energy scales,  which helps `screen' the Higgs mass from large loop corrections, while the flavour structure emerges essentially from a `deconstruction' of color and hypercharge.

The paper is organised as follows. In Section~\ref{sec:assumptions} we state our 
model-building hypotheses and general strategy to identify what we regard as viable models. 
In Section~\ref{sec:bottom-up} we exploit these hypotheses to classify all the relevant 
TeV-scale gauge groups. In Section~\ref{sec:mixing}  we analyse the main 
mechanism of flavour mixing in these models, deriving further constraints on their matter content.
In Section~\ref{sec:UV} we outline  general ideas about the UV 
completions of the models, especially concerning the dynamics responsible for resolving the light fermion families.
In Section~\ref{sec:UVwinning} we present a more detailed discussion of the UV completion 
of a specific model, not considered in previous literature, which emerges as the 
most successful in satisfying all our naturalness requirements.
The results are summarised in the Conclusions.

\label{sec:mixing}

\section{Main model-building hypotheses and general strategy} 
\label{sec:assumptions}

The  main hypotheses that guide our model-building analysis can be summarised as follows.
\begin{enumerate}
\item {\bf Flavour non-universality.} We assume that the approximate flavour symmetries observed in the SM Yukawa couplings are 
accidental symmetries resulting from an intrinsically flavour non-universal gauge group in the UV.  The flavour non-universality of gauge interactions
can be manifest at different energy scales; however, the stability of the Higgs sector (point~3) requires that part of it  -- the separation between the third generation and the light families --
occurs already at the TeV scale.  
\item {\bf Semi-simple embedding in the UV.} 
We assume the TeV scale model eventually embeds into an anomaly-free semi-simple gauge theory, explaining hypercharge quantization. This final embedding in the UV model does not need to occur at a low-scale, but we only consider models which have such a UV embedding.\footnote{The space of (potentially flavour non-universal) semi-simple completions of the SM, including up to three right-handed neutrinos but no further chiral fermions, was completely categorised in Ref.~\cite{Allanach:2021bfe}. To exploit these results, and to limit the scope of our study, we make the same assumption that there are no {\em chiral} fermions beyond the SM+$3\nu_R$ -- we do, however, allow for the possibility of vector-like fermions.  }
\item {\bf Fundamental and (quasi)-natural Higgs at the TeV scale.} We assume the Higgs is a fundamental scalar field up to scales well above the TeV.
We do not exclude UV completions with supersymmetry or Higgs compositeness. However, we assume that these mechanisms, if utilised, enter at scales 
suitably above the first layer of new physics that involves the third generation (motivated by points 1 and 2 above). The request that the Higgs mass corrections are not too large is then used as a criterion to constrain the structure and spectrum of the viable models. 
\item {\bf Avoiding proton decay and $\mu\to e$ at low scales.} Among the unification patterns consistent with point~2, we disregard those leading to
 proton decay and lepton-flavor violating processes of the type $\mu\to e$ at low energy scales. 
\end{enumerate}
As we shall see in the next two Sections, these guiding principles lead us to identify a very constrained set of viable models at the TeV scale, with 
interesting common phenomenological features. The same principles allow us also to draw interesting conclusions concerning their possible 
UV completions, that we discuss in general terms in Section~\ref{sec:UV}, and in more detail for a particularly motivated case in Section~\ref{sec:UVwinning}.

\section{Flavour non-universality at the TeV scale}
 \label{sec:bottom-up}
 
The SM is an anomaly-free chiral gauge theory with gauge group\footnote{Throughout this paper we use a convention for labelling groups in which $G$ (with possible subscripts) is reserved for semi-simple Lie groups, while a group labelled by
$H$ (with possible subscripts) will be reductive but not semi-simple -- it will contain $\u$ factors. \label{footnote:semisimple}
}
\footnote{The group written in Eq.~(\ref{eq:SM}) is not the only possibility for the SM gauge group; we have ignored the possibility of a discrete quotient by a normal subgroup isomorphic to $\Z_2$, $\Z_3$, or $\Z_6$ (see {\em e.g.}~\cite{Tong:2017oea}). Such discrete quotients, both in the SM gauge symmetry and in the various BSM gauge groups we consider, will play no role in our discussion.  }
\be \label{eq:SM}
\sm := \su(3) \product \su(2)_L \product \u_Y\, .
\ee
The Higgs transforms as $H \sim \left( {\bf 1, 2,} -3 \right)$,
where we use the minimal integral normalisation for hypercharge in which the left-handed quark doublet has charge $1$. Including three SM singlet right-handed neutrinos, there are $24$ left-handed and $24$ right-handed Weyl fermions, which we denote schematically by $\psi_L$ and $\psi_R$.
In the SM, the family-universal gauge interactions allow for Yukawa couplings that give mass to every fermion, of the form
\be
Y_{ij} \, \overline{\psi}_L^i H^{(c)} \psi_{R}^j\,\, ,
\ee
where $i, j$ are family indices, and $H^c = i\sigma_2 H^\dagger \sim \left( {\bf 1, 2,} +3 \right)$.

Our scope in this Section is to identify flavour non-universal gauge groups, obtained by deconstructing $\sm$ in flavour space, 
that would justify the leading hierarchical pattern exhibited by the $Y_{ij}$.
With this purpose, it is useful to consider more closely the action of the different SM gauge factors on fermions and the Higgs.
The strong interaction is vector-like, and moreover all leptons and the SM Higgs are neutral under it.
The $\su(2)_L$ interaction acts only on $\psi_L$ and the Higgs. Hypercharge acts chirally on all SM fermion fields and non-trivially on the Higgs, and can be decomposed as a linear combination of a vector-like part ($B-L$, under which $H$ is neutral) and a purely right-handed part ($T^3_R$, under which $H$ has charge $-1/2$).  The SM field representations under the $\u$-extended $\su(3) \product \su(2)_L \product \u_{B-L} \product \u_R$ symmetry are summarised in Table~\ref{tab:SMfermions}.

\begin{table}
\begin{center}
\begin{tabular}{|c|c||c|ccc|}
\hline
Field(s) & Chirality & $\su(3)$ & $\su(2)_L$ & $\u_{B-L}$ & $\u_R$ \\
\hline
$H$ & - & ${\bf 1}$ & ${\bf 2}$ & $0$ & $-3$ \\
$\psi_L$ & L & ${\bf 3}$ or ${\bf 1}$ & ${\bf 2}$ & $-3$ or $1$ & $0$\\
$\psi_R$ & R & ${\bf 3}$ or ${\bf 1}$ & ${\bf 1}$ & $-3$ or $1$ & $\pm 3$\\
\hline
\end{tabular}
\end{center}
\caption{Representations of SM fields under $\su(3) \product \su(2)_L \product \u_{B-L} \product \u_R$.
} \label{tab:SMfermions}
\end{table}

\subsection{Accidental flavour symmetries from gauge non-universality}
  The leading hierarchical pattern exhibited by the $Y_{ij}$ is well described by an approximate $\UU(2)^5$ symmetry, that acts on each pair of light fermion fields in the appropriate doublet representation, and on each third generation fermion as a singlet.
In view of this, we consider the implications of separately deconstructing each of the 
gauge factors in Table~\ref{tab:SMfermions} into independent symmetries acting on the light {\em vs.} the third generation. 
The Higgs quantum numbers are set to allow its coupling to the third generation. The renormalisable Yukawa couplings $Y_{ij}$ permitted by the different options can be summarised as follows.
\begin{itemize}[leftmargin=*]
\item Deconstructed $\su(3)^\light \product \su(3)^\heavy$ allows
\be
Y^F_{ij} \sim \begin{pmatrix}
\times&\times&0\\
\times&\times&0\\
0&0&\times
\end{pmatrix}\, , \qquad F \in \{U, D\}\, ,
\ee
but it gives no constraint on the lepton Yukawa.
\item Deconstructed $\su(2)_L^\light \product \su(2)_L^\heavy$, with Higgs charged as ${\bf (1, 2)}$, allows
\be
Y^F_{ij} \sim \begin{pmatrix}
0&0&0\\
0&0&0\\
\times&\times&\times
\end{pmatrix}\, , \qquad F \in \{U, D, E\}\, .
\ee
\item Deconstructed $\u_{B-L}^\light \product \u_{B-L}^\heavy$ allows
\be \label{eq:YukBminusL}
Y^F_{ij} \sim \begin{pmatrix}
\times&\times&0\\
\times&\times&0\\
0&0&\times
\end{pmatrix}\, , \qquad F \in \{U, D, E\}\, .
\ee
\item Deconstructed $\u_R^\light \product \u_R^\heavy$, with Higgs charged as $(0,-3)$, allows
\be
Y^F_{ij} \sim \begin{pmatrix}
0&0&\times\\
0&0&\times\\
0&0&\times
\end{pmatrix}\, , \qquad F \in \{U, D, E\}\, .
\ee
\end{itemize} 
Unsurprisingly, deconstructing the $\su(2)_L$ or the $\u_R$ interaction distinguishes the third generation from the light ones, because the Higgs is charged under these symmetries; however,  neither on its own is enough to allow only the $Y_{33}$ couplings. Even though the Higgs is neutral under $B-L$, deconstructing $B-L$ bans Yukawas that {\em mix} the light fermions with the third generation. Deconstructing $\su(3)$ would have a similar implication but only for the quarks, giving no constraint on the charged lepton Yukawa matrix. 

The upshot is that, if we deconstruct {\em any pair} of $\su(2)_L$, $\u_{B-L}$, and $\u_R$, then we allow only  the $Y_{33}$ entries of the Yukawa couplings for both quarks and charged leptons, and thereby deliver the $U(2)^5$ accidental symmetry.\footnote{In a similar spirit, flavour non-universal hypercharge extensions of the SM, designed to only allow the $Y_{33}$ couplings while also making a connection to possible new physics in $b\to s\ell\ell$ transitions, were suggested in Refs.~\cite{Allanach:2018lvl,Davighi:2019jwf,Allanach:2019iiy,Davighi:2021oel}. }
This suggests three different paths for the `next layer' of new physics, which we would like to enter at the TeV scale on grounds of naturalness. 
Namely, the embedding of $\sm$ inside the TeV-scale group $H_{\text{TeV}}$ should factor through one of the following three (reductive) subgroups:
\begin{align}
H_{\text{TeV}} &\supset \su(3) \product \su(2)_L \product {\u_{B-L}^\light} \product {\u_R^\light} \product {\u_{B-L}^\heavy}  \product {\u_R^\heavy}\, , \label{eq:gTEV1} \\
H_{\text{TeV}} &\supset \su(3) \product \u_{B-L} \product {\su(2)_L^\light} \product {\u_R^\light}  \product {\su(2)_L^\heavy} \product {\u_R^\heavy}\, ,  \label{eq:gTEV2} \\
H_{\text{TeV}} &\supset \su(3) \product \u_R \product {\su(2)_L^\light} \product {\u_{B-L}^\light}  \product {\su(2)_L^\heavy}  \product {\u_{B-L}^\heavy} \, . \label{eq:gTEV3}
\end{align}
We can of course also consider deconstructing all three of $\su(2)_L$, $\u_{B-L}$, and $\u_R$ at the TeV scale, arriving at the less-minimal option:
\begin{align} \label{eq:gTEV4}
\su(3) \product {\su(2)_L^\light}  \product {\u_{B-L}^\light} \product {\u_R^\light} \product {\su(2)_L^\heavy} \product {\u_{B-L}^\heavy} \product {\u_R^\heavy}\, , 
\end{align}
which contains each of (\ref{eq:gTEV1}--\ref{eq:gTEV3}) as subgroups.
All these gauge symmetries take the form
\be 
\h_U \product {\h_{12}} \product {\h_3} \, ,
\ee
where the Higgs is charged under $\h_3$ and possibly $\h_U$, and all are anomaly-free.

\subsection{Third family quark-lepton unification}

As stated in Section~\ref{sec:assumptions}, one of our key requests is to ultimately embed the gauge model in an anomaly-free semi-simple UV completion. 
All $\u$ factors in the gauge symmetry must eventually be `absorbed' into non-abelian (semi-simple) factors at some energy scale.

For the non-universal gauge groups (\ref{eq:gTEV1}--\ref{eq:gTEV4}), the Higgs and the third generation fermions $\psi^3$ (to which the Higgs is strongly coupled) are charged only under $\h_U \product \h_3$. These groups always include  $\u_R$ and $\u_{B-L}$ factors, whose embedding into a semi-simple factor would imply 
exotic heavy gauge bosons at the corresponding unification scale. Since these new states couple directly to the Higgs and/or $\psi^3$, 
our (quasi)-naturalness assumption likely requires this unification layer to occur around the TeV scale, similarly to the breaking of the deconstructed flavour groups (\ref{eq:gTEV1}--\ref{eq:gTEV4}) 
into $\sm$.

An important consequence is that the embedding of $\u_{B-L}^\heavy$ inside a simple factor necessarily implies quark-lepton unification, and so produces vector leptoquarks that couple to the third family.
As we discuss below, this significantly restricts the viable options given the constraint from proton stability and, to a lesser extent, 
$\mu\to e$ transitions. 

\subsection{Naturalness of the extended gauge sector}
\label{sect:NatG}

We here provide a more quantitative evaluation of the natural scale of the first unification and flavour-deconstruction layers by estimating the 
quantum corrections to the Higgs mass induced by the heavy gauge bosons. 

Unifying the electroweak factors of $\su(2)_L$, $\su(2)_L^\heavy$, $\u_R$, or $\u_R^\heavy$ into larger symmetries, as well as breaking the deconstructed $\su(2)_{L/R}^\light\product \su(2)_{L/R}^\heavy$ down to the SM, produces (flavoured) electroweak gauge bosons $\{X\}$ that couple directly to the Higgs. From these new degrees of freedom
we expect one-loop corrections of order
\be \label{eq:EWboson_correction}
\delta m_h^2 \sim \frac{1}{16\pi^2}\, g_{L/R}^2\, M_X^2\, ,
\ee
where $M_X$ is the gauge boson mass, and the gauge coupling $g_{L/R}$ can be estimated by the values of the SM gauge couplings $\su(2)_L/\u_Y$ (at the TeV scale). Employing the finite naturalness condition~\cite{Farina:2013mla}
\be 
\frac{|\delta m_h^2|}{m_h^2} \lesssim \Delta \,,
\label{eq:naturalness}
\ee
where $\Delta$ quantifies the degree of fine-tuning, implies $M_{X} \lesssim$~a few TeV for $\Delta \approx 1$.\footnote{This naturalness estimate will be refined in \S \ref{sec:VLF-gauge-interplay} when we consider the generation of light-quark Yukawa couplings; for now, it should be understood as a very rough guideline.}
We take this as a guideline limit on the scale of $\su(2)_{L/R}^\light\product \su(2)_{L/R}^\heavy$ breaking, and the scale of any further unification of $\su(2)_L$, $\su(2)_L^\heavy$, $\u_R$, or $\u_R^\heavy$ factors. This is summarised in the first row of Table~\ref{tab:loop-order}.

\begin{table}
\begin{center}
\begin{tabular}{|c||c|c|}
\hline
Symmetry & $\delta m_h^2$ order & Natural scale \\
\hline
$\su(2)_L$,  $\u_R$, $\su(2)_L^\heavy$, $\u_R^\heavy$ & 1 loop & $\sim$ TeV \\
$\su(3)$,  $\u_{B-L}$, $\u_{B-L}^\heavy$ & 2 loops & $\lesssim 10$ TeV \\
\hline
\hline
All $\h_{12}$ factors & $\geq 2$ loops & $\gg 10$ TeV \\
\hline
\end{tabular}
\end{center}
\caption{Loop order of Higgs-mass corrections and corresponding natural scale for the exotic gauge bosons.
The symmetry factors of interest, that appear in the  non-universal gauge groups (\ref{eq:gTEV1}--\ref{eq:gTEV4}), are listed in the first column.
The second column indicates at which loop order an associated exotic gauge boson affects the Higgs mass. 
In the third column we report the upper limit on the scale at which the symmetry group can be unified into a larger symmetry
according to the Higgs-mass constraint in (\ref{eq:naturalness}).  The gauge bosons in $\h_{12}$ can talk to the Higgs at the two-loop order; however, as we shall discuss 
in \S\,\ref{sec:UV},  their contributions to the Higgs mass can be further parametrically suppressed by {\em e.g.}~the light fermion Yukawa couplings. 
 \label{tab:loop-order} }
\end{table}

The second row of Table~\ref{tab:loop-order} lists symmetries which do not couple directly to the Higgs, but which do couple directly to the top quark. Their breaking does not result in one-loop Higgs mass corrections, but rather the quantum correction starts at two-loops via diagrams with gauge bosons (leptoquarks and/or $\ZP$ bosons) running inside a top quark loop. 
Taking as an example the contribution from a $U_1$ leptoquark associated with $\su(3)\product \u_{B-L} \hookrightarrow \su(4)$ quark-lepton unification, the finite Higgs mass correction resulting from these diagrams scales as
\be \label{eq:LQ_correction}
\delta m_h^2 \sim \left( \frac{1}{16\pi^2} \right)^2 \,N_c \,y_t^2\, g_s^2\, M_{U}^2\, ,
\ee
where $N_c$ is a colour factor, $y_t$ is the top quark Yukawa coupling, and $g_s$ is an $\mathcal{O}(1)$ gauge coupling which matches onto the strong coupling at the scale of symmetry breaking. 
Note that the loop factor here appears squared, in contrast to (\ref{eq:EWboson_correction}). Proceeding as above we deduce
$M_{U} \lesssim 10~\text{TeV}$ for $\Delta \approx 1$, and we take this as upper scale scale for $\u_{B-L}^\heavy$  deconstruction and/or third family quark-lepton unification.
Note that this scale is not a whole loop factor higher than the scale associated with the electroweak deconstruction/unification. This is  because the Higgs mass corrections,
while being generated beyond the one-loop level, necessarily involve large couplings ($y_t$, the top quark Yukawa coupling, and $g_s > g_{L/R}$, the strong gauge coupling). These naturalness-derived guidelines are recorded in Table~\ref{tab:loop-order}.

The first two lines in Table~\ref{tab:loop-order} indicate rather low scales,  implying interesting phenomenology within the reach of existing facilities. 
Of course, the effects of the associated gauge bosons must be out of reach of present collider and flavour bounds, which means that not all the algebraic possibilities are phenomenologically viable at the unification scales dictated by naturalness. This helps one to narrow down the options significantly, as we discuss next.

\subsection{Constraints from proton stability and $\mu\to e$ transitions}
\label{sec:proton}

While factors of $\u_R$ can be absorbed into $\su(2)_R$ factors near the TeV scale with moderate tuning to respect phenomenological bounds, 
the low-scale leptoquarks resulting from the third family quark-lepton unification requires more care.

Restricting the attention to a single generation (or to `vertical' embeddings that treat multiple families universally),  the basic unification patterns  
which do not require extra fermions use 
$\su(4)$ \`a la Pati and Salam~\cite{Pati:1974yy}, or $\su(5)$ \`a la Georgi and Glashow~\cite{Georgi:1974sy} 
(hence, $\so(10)$~\cite{georgi1975particles,Fritzsch:1974nn} also). The $\su(5)$ option cannot be used near the TeV scale because of the very stringent constraint coming from the non-observation of decaying protons. This is true both for the  flavour-universal $\u_{B-L}$ but  also for $\u_{B-L}^\heavy$ (due to constraints on baryon-number violating decays of bottom-baryons, as well as those from proton decay after the CKM mixing propagates the baryon-number violating effects into the light generations).
As a result,  the request of incorporating all $\u$ factors in $H_U \product H_3$ into simple factors around the $1-10$ TeV scale leaves open only the Pati--Salam 
option. In view of the above considerations, in Table~\ref{tab:TeV} we show the possible anomaly-free\footnote{It is straightforward to check that all perturbative anomalies, as well as potential non-perturbative anomalies associated with the various $\su(2)$ factors~\cite{Witten:1982fp}, cancel. This is enough to guarantee full non-perturbative anomaly cancellation for these theories, as can be checked by computing bordism groups via similar methods to Refs.~\cite{Garcia-Etxebarria:2018ajm,Davighi:2019rcd}.
} gauge groups satisfying the assumptions listed in Section~\ref{sec:assumptions}
 for the first unification/flavour-deconstruction layer at a scale of a few TeV. 

\definecolor{DarkGray}{gray}{0.8}
\definecolor{LightGray}{gray}{0.92}
\begin{table}
\hglue -0.7 cm
\begin{tabular}{|c|c|c|c|c|}
\hline
\multicolumn{5}{|c|}{$\g_U \product \g_3 \product H_{12}$} \\
\hline
\hline
& $\g_U$ & $\g_{3}$ & $H_{12}$ & Example \\
\hline
1& $\su(2)_L\!$ & $\su(4)^\heavy \product \su(2)_R^\heavy$ & $\su(3)^\light \product \u_{B-L}^\light \product \u_R^\light$ & \S~\ref{sec:UVwinning} \\ \hline
\rowcolor{LightGray} 
2& $\su(2)_R\!$ & $\su(4)^\heavy \product \su(2)_L^\heavy$ & $\su(3)^\light \product \su(2)_{L}^\light \product \u_{B-L}^\light$ & $\times$ \\ \hline
\rowcolor{DarkGray} 
3& $\su(4)$ & $\su(2)_L^\heavy \product \su(2)_R^\heavy$ & $\su(2)_L^\light \product \u_R^\light$ & \cite{Davighi:2022fer} \\ 
\hline
4& $\emptyset$ & $\su(4)^\heavy \productn \su(2)_L^\heavy \productn \su(2)_R^\heavy$ & $\su(3)^\light \productn \su(2)_{L}^\light \productn \u_{B-L}^\light \productn \u_R^\light$ & \cite{Bordone:2017bld,Fuentes-Martin:2022xnb,Davighi:2022bqf} \\
\hline
\end{tabular}
\caption{Candidate gauge symmetries for the first unification/deconstruction layer at a scale of a few TeV.  
The third row is darkly shaded because of the O($100$~TeV) constraint on the mass of the (flavour-universal) leptoquark from $K_L \to \bar{e}\mu$.
The second row is lightly-shaded because of the excessive right-handed mixing between light- and third-generation fermions (\S\,\ref{sec:mixing}).
 \label{tab:TeV} }
\end{table}

Finally, the color coding in Table~\ref{tab:TeV} indicates the viability of each option at its natural scale, according to general bounds from flavour-violating processes. 
A severe constraint follows from the non-observation of  $K_L \to \bar{e}\mu$, which implies that a $U_1$ leptoquark with $O(1)$ couplings to the light families should have mass greater than $200$ TeV~\cite{Giudice:2014tma},  in sharp conflict with the naturalness assumption (\ref{eq:naturalness}). This  disfavours option~3 in Table~\ref{tab:TeV}. 
As we discuss in Section~\ref{sec:mixing}, option~2 is also disfavoured (although less severely) by the unsuppressed right-handed mixing between 
light and third generation fermions which, in the absence of tuning, would yield excessively large right-handed flavour-changing currents. We are thus left with two particularly 
motivated options for TeV scale dynamics: options 1 and 4. The latter has already been discussed in the literature, with possible UV 
completions proposed in~\cite{Bordone:2017bld,Fuentes-Martin:2022xnb,Davighi:2022bqf}. A UV completion for the option~1, not discussed so far in the literature, 
is presented in Section~\ref{sec:UVwinning} of the present paper.

We emphasise that models based on the gauge groups 2 and 3 in Table~\ref{tab:TeV} are certainly not ruled out experimentally; rather, satisfying phenomenological bounds for these models leads to a conflict with our assumptions listed in Section~\ref{sec:assumptions}. In particular, more tuning on the Higgs mass and/or the flavour couplings is required. A notable example in this class is the model of electroweak-flavour-unification~\cite{Davighi:2022fer,Davighi:2022vpl} in all three families, based on the gauge group $\su(4) \product \sp(6)_L \product \sp(6)_R$. In such a model, the flavour bounds on the associated vector leptoquark imply large Higgs-mass corrections coming from the $\su(4)$ symmetry breaking scale 
(in the absence of further ingredients, such as supersymmetry or compositeness).

\section{Mixing between the light and heavy generations} \label{sec:mixing}

For all the models we consider, which admit a factorization {\em viz.} $\g_U \product \g_3 \product H_{12}$, there must exist extra fields that transform under both $H_{12}$ and $\g_3$, in order to mix the light fermions with $\psi^3$, and to generate Yukawa couplings for the light fermions. Continuing with our bottom-up analysis, we here consider the simplest possibilities for this mixing, starting from an effective field theory (EFT) approach.

\subsection{EFT operators} \label{sec:EFTmix}

The breaking of the flavour-deconstructed groups in Table~\ref{tab:TeV}  into $\sm$ can be minimally achieved by a set of bifundamental scalar fields acquiring non-vanishing vacuum expectation values (vevs):
\begin{align}
&\Omega_3 \sim ({\bf \overline{4}, 3}, 1)\,, \quad \Omega_1 \sim ({\bf \overline{4}, 1}, -3)
	&&\text{of~~} \su(4)^\heavy \product \su(3)^\light \product \u_{B-L}^\light \,, \label{eq:Om3} \\
&\Sigma_L \sim ({\bf 2, 2}) 
	&&\text{of~~} \su(2)_L^\heavy \product \su(2)_L^\light\,, \label{eq:SigmaL} \\
&\Sigma_{R\pm} \sim ({\bf 2}, \pm 3)
	&&\text{of~~} \su(2)_R^\heavy \product \u_R^\light\,, \label{eq:SigmaR-}
\end{align}
where each field transforms trivially under all remaining gauge factors not written. 
For example, particular non-zero vevs for $\Omega_3$ and $\Omega_1$ trigger a breaking $\su(4)^\heavy \product \su(3)^\light \product \u_{B-L}^\light \to \su(3) \product \u_{B-L}$, as is typical in 4-3-2-1 models~\cite{DiLuzio:2017vat}.

In an EFT approach,  we now analyse the Yukawa textures obtained by combining the SM fermions, the Higgs field, and the vevs of these scalar link fields, into a tower of effective operators respecting the gauge symmetries in Table~\ref{tab:TeV}.

\subsection*{Model 1}

For the model in row 1 of Table~\ref{tab:TeV}, with flavour-universal $\su(2)_L$ symmetry and a single Higgs field charged in the $({\bf 2, 2})$ representation of $\su(2)_L \product \su(2)_R^\heavy$, the leading effective Yukawa interactions involving light fields arise from
\be \label{eq:EFTmodel1}
\cL_{\leq 6} \supset \frac{c_{i3}}{\Lambda_\Omega}\, \overline{\psi}^i_{L} \Omega_{3(1)} H^{(c)} \psi^3_{R} + \frac{c_{3i}}{\Lambda_\Omega \Lambda_R }\, \overline{\psi}^3_{L} \Omega_{3(1)} H^{(c)}\Sigma_{R\pm}  \psi^i_{R} + \frac{c_{ij}}{\Lambda_R}\, \overline{\psi}^i_{L}  H^{(c)} \Sigma_{R\pm}\psi^j_{R} \, ,
\ee
where $i,\, j \in \{1,2\}$ are here light-family indices. The $c_{ik}$ are Wilson coefficients that, in absence of an underlying flavour structure distinguishing the two 
light families, are expected to be $\mathcal{O}(1)$.
Let us express the vevs of the link fields as 
\begin{align}
v_4:= |\langle \Omega_{3(1)} \rangle| \sim \epsilon_{\Omega} \Lambda_\Omega, \qquad 
v_L:= |\langle \Sigma_{L} \rangle| \sim \epsilon_{L} \Lambda_L, \qquad 
v_R:= |\langle \Sigma_{R\pm} \rangle| \sim \epsilon_{R} \Lambda_R,
\end{align}
where $|\langle S \rangle|$ denotes the magnitude of the vev of a scalar field $S$ in a non-singlet representation,
and where the (smallest) heavy scale $\Lambda$ indicates the cut-off of the EFT.
We obtain the following Yukawa texture:
\be
\text{Model 1:} \qquad
Y^F \sim \begin{pmatrix}
\epsilon_R & \epsilon_\Omega \\
\epsilon_\Omega \epsilon_R & 1
\end{pmatrix}\, , 
\qquad F\in\{U,D,E\}\, .
\label{eq:text1}
\ee
Here $Y^F$ is written as a block matrix reflecting the 2+1 structure: the upper-left element implicitly denotes a $2\times2$ matrix, the upper-right element a 2-component column vector, 
and the lower-left element a 2-component row vector.

The texture (\ref{eq:text1}) is a very economical and efficient way  for modelling the observed pattern of the Yukawa couplings. 
We recover without any tuning the desired leading (and minimal) breaking of the $\UU(2)^5$ flavor symmetry~\cite{Barbieri:2011ci,Isidori:2012ts}, setting 
\be
\epsilon_\Omega \sim |V_{cb}|, \qquad \epsilon_R \sim \frac{y_2}{y_3}\, .
\ee
The range implied by quark/lepton masses and quark mixing for these parameter is $10^{-2}-10^{-1}$.
Of course at this level we are not able to address the splitting among light families, that we assume arises from higher energy scales (see \S\,\ref{sec:UV}).
An important point to stress is that we naturally obtain the hierarchy
\be
|Y^F_{3i}| \ll |Y^F_{i3}|\, ,
\ee
which ensures suppressed right-handed mixing between light and heavy fields. 
This is a highly desirable feature, phenomenologically, for flavoured new physics near the TeV scale.

\subsection*{Model 2}

For the model in row 2 of Table~\ref{tab:TeV}, with flavour-universal $\su(2)_R$ symmetry and a single Higgs field charged in the $({\bf 2, 2})$ representation of $\su(2)_L^\heavy \product \su(2)_R$, the effective Yukawa interaction arises from
\be \label{eq:EFTmodel2}
\cL_{\leq 6} \supset \frac{c_{i3}}{\Lambda_\Omega \Lambda_L}\, \overline{\psi}^i_{L,} \Sigma_L \Omega_{3(1)} H^{(c)} \psi^3_{R} + \frac{c_{3i}}{\Lambda_\Omega}\, \overline{\psi}^3_{L} \Omega_{3(1)} H^{(c)} \psi^i_{R} + \frac{c_{ij}}{\Lambda_L}\, \overline{\psi}^i_{L} \Sigma_{L} H^{(c)} \psi^j_{R}  \, .
\ee
The third column of the Yukawa matrices are now populated at dimension-6, while the third row is populated at dimension-5, yielding:
\be
\text{Model 2:} \qquad
Y^F \sim \begin{pmatrix}
\epsilon_L & \epsilon_\Omega \epsilon_L \\
\epsilon_\Omega & 1
\end{pmatrix}\, , 
\qquad F\in\{U,D,E\}\, .
\ee
This texture implies a  mixing in the right-handed sector naturally larger than that in the left-handed one, which is in conflict with observations.
More precisely, the need to satisfy the strong phenomenological constraints on right-handed flavour-changing currents (see {\em e.g.}~\cite{Isidori:2014rba,LHCb:2020dof}) and, 
at the same time, reproduce the observed pattern of Yukawa eigenvalues and left-handed mixing (giving rise to the CKM matrix), are in conflict. 
These conditions can here be satisfied only via a severe tuning of the coefficients in 
the effective Lagrangian~(\ref{eq:EFTmodel2}). We conclude that model 2 does not yield a natural flavour model.

\subsection*{Models 3 and 4}

Proceeding in a similar manner, one arrives to the following similar textures for the remaining models in Table~\ref{tab:TeV}:
\begin{eqnarray}
&\text{Model 3:} \qquad
&Y^F \sim \begin{pmatrix}
\epsilon_L \epsilon_R & \epsilon_L \\
\epsilon_R & 1
\end{pmatrix}\, , 
\qquad\quad F\in\{U,D,E\}\, . \\
&\text{Model 4:} \qquad
&Y^F \sim \begin{pmatrix}
\epsilon_L \epsilon_R  & \epsilon_{\Omega} \epsilon_L \\
\epsilon_{\Omega} \epsilon_R & 1
\end{pmatrix}\, , 
\qquad  F\in\{U,D,E\}\, .
\end {eqnarray}
In both cases a sufficient suppression of right-handed currents is achieved imposing
\be \label{eq:eps_hierarchy}
\epsilon_R \ll \epsilon_L\, .
\ee
which is a natural possibility.\footnote{In an explicit UV model it may not even be necessary to invoke such a hierarchy of vevs; depending on the specific representations of the underlying VLFs, it is possible that the $Y_{3i}^F$ elements of the Yukawa matrices are not generated at all at tree level. This is the case, for example, in the model of Ref.~\cite{Davighi:2022bqf}.} 

Focusing on model 4, which is the only other option consistent with our model-building hypotheses (\S\ref{sec:proton}), one can obtain a realistic pattern for a wide range of  $\epsilon_{\Omega}$,
even up to $\epsilon_{\Omega}\sim \mathcal{O}(1)$, which is the case relevant to 4-3-2-1 models addressing the $B$-physics 
anomalies~\cite{Bordone:2017bld, Greljo:2018tuh,Fuentes-Martin:2019ign,Fuentes-Martin:2020hvc,Fuentes-Martin:2020bnh,Fuentes-Martin:2022xnb,Davighi:2022bqf}. 
However, it is fair to say that the most motivated pattern, considering only the observed patter of the Yukawa couplings and all the existing bounds on 
flavour-changing processes, is the model~1 structure in (\ref{eq:text1}).

\subsection{Completion via vector-like fermions} \label{sec:VLFcompletion}

For all the four classes of TeV-scale models we have identified, a simple UV origin for the effective operators in \S\,\ref{sec:EFTmix} is via vector-like fermions (VLFs) with mass $\mathcal{O}(\Lambda)$. Below we provide an explicit mechanism in the case of model 1, which we have identified as the simplest and most natural option, also noting that models in this class have not been studied before in the literature to our knowledge.

\subsection*{VLFs for model 1}

\begin{figure*}[t]
\begin{center}
\begin{tikzpicture}
\begin{feynman}
\vertex (a) { $\overline{\psi}_L^i$};
\vertex [right=0.6in of a] (b);
\vertex [right=0.4in of b] (c);
\vertex [right=0.4in of c] (d);
\vertex [right=0.4in of d] (e) { $\psi_R^3$};
\node at (b) [circle,fill,inner sep=1.5pt,label=below:{ $\,\,\,\, y_{\lambda \Omega}$}]{};
\node at (d) [circle,fill,inner sep=1.5pt,label=below:{ $y_{\lambda H} \,\,$}]{};
\node at (c) [square dot,fill,inner sep=1.0pt]{};
\vertex [above=0.5in of b] (f) { $\Omega_{3(1)}$};
\vertex [above=0.5in of d] (g) { $H^{(c)}$};
\diagram* {
(a) -- [plain] (b)  -- [plain,edge label={ {\footnotesize$\lambda_R$}}] (c) -- [plain,edge label={ {\footnotesize$\overline{\lambda}_L$}}] (d) -- [plain] (e),
(b) -- [scalar] (f),
(d) -- [scalar] (g),
};
\vertex [right=0.6in of e] (a1) { $\overline{\psi}_L^i$};
\vertex [right=0.6in of a1] (b1);
\vertex [right=0.4in of b1] (c1);
\vertex [right=0.4in of c1] (d1);
\vertex [right=0.4in of d1] (e1) { $\psi_{R}^j$};
\node at (b1) [circle,fill,inner sep=1.5pt,label=below:{ $\,\,y_{\rho H}$}]{};
\node at (c1) [square dot,fill,inner sep=1.0pt]{};
\node at (d1) [circle,fill,inner sep=1.5pt,label=below:{ $y_{\rho \Sigma}\,\,$}]{};
\vertex [above=0.5in of b1] (h1) { $H^{(c)}$};
\vertex [above=0.5in of d1] (i1) { $\Sigma_{R\pm}$};
\diagram* {
(a1) -- [plain] (b1)  -- [plain,edge label={ {\footnotesize $\rho_R$}}] (c1) -- [plain,edge label={ {\footnotesize $\overline{\rho}_L$}}] (d1) -- [plain] (e1),
(b1) -- [scalar] (h1),
(d1) -- [scalar] (i1),
};
\end{feynman}
\end{tikzpicture}
\end{center}
\caption{Tree-level Feynman diagrams generating effective Yukawa couplings that mix the light left-handed fermions with third generation right-handed fermions (left diagram), and the light-light SM Yukawa couplings (right diagram), in model 1 (see Table~\ref{tab:TeV}). \label{fig:VLFmodel1a} }    
\end{figure*}
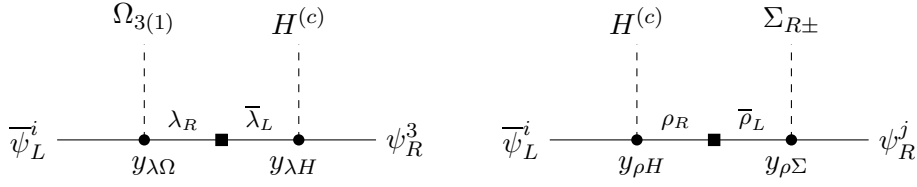

Consider extending the model by a pair of VLFs in the representations
\begin{align}
&\lambda_{L/R} \sim ({\bf {4}, 2})    &&\text{of~~} \su(4)^\heavy \product \su(2)_L\, , \label{eq:lambda} \\
&\rho_{L/R} \sim ({\bf {3}, 2}) \oplus ({\bf {1}, 2})   &&\text{of~~} \su(3)^\light \product \su(2)_R^\heavy\, , \label{eq:rho}
\end{align}
including the following renormalisable interactions in the Lagrangian
\begin{align}
\cL \supset \,\, &M_\lambda\, \overline{\lambda}_L \lambda_R + M_\rho\, \overline{\rho}_L \rho_R \\
+ &y_{\lambda H}\, \overline{\lambda}_L H^{(c)} \psi_R^3 + y_{\lambda \Omega}^i\, \overline{\psi}_L^i \Omega_{3(1)} \lambda_R \label{eq:ylambdaH}\\
+ &y_{\rho H}^i \, \overline{\psi}_L^i H^{(c)} \rho_{R} + y_{\rho \Sigma}\, \overline{\rho}_{L} \Sigma_{R\pm} \psi_R^i\, , \qquad i\in\{1,2\}\, . \label{eq:yrhoH}
\end{align}
With these interactions, after integrating out both the VLFs one generates the following Yukawa couplings in Eq.~(\ref{eq:EFTmodel1}):
\begin{itemize}
\item The dimension-5 operators occupying the third column of the Yukawa matrices, after integrating out $\lambda$.
\item The dimension-5 operators occupying the upper-left 2-by-2 block, {\em i.e.} the light-light Yukawa couplings, after integrating out $\rho$.
\end{itemize}
The tree-level Feynman diagrams are shown in Fig.~\ref{fig:VLFmodel1a}. The heavy scale associated with $|V_{cb}|$ is therefore $\Lambda_{\Omega} := M_\lambda/{y_{\lambda \Omega} y_{\lambda H}}$, such that $\epsilon_\Omega = v_\Omega/ \Lambda_{\Omega} \sim |V_{cb}|$; the heavy scale associated with the hierarchy $y_2/y_3$ is   $\Lambda_R := M_\rho/{y_{\rho H} y_{\rho \Sigma}}$, such that $\epsilon_R = v_R/\Lambda_R \sim y_2/y_3$.

Introducing only $\lambda$ and $\rho$, the dimension-6 operators that populate the third row of the Yukawa matrix in Eq.~(\ref{eq:EFTmodel1}) are not in fact generated at tree-level. One can see this simply from the fact that neither VLF has a coupling to $\psi_{L}^3$. 
Such a scenario is phemenologically viable, and is indeed preferred by data (for the TeV scale new physics that we consider here), since it strongly suppresses right-handed flavour-changing currents. We therefore learn that, going beyond the EFT and considering explicit completions via VLFs, it is very natural in minimal extensions for these subleading Yukawa terms to not be generated at tree-level at all.

For completeness, we note that these third-row Yukawa operators could, however, be generated at tree-level by including a third VLF,  
$\tilde\rho$, transforming in the representation $({\bf \overline{4}, 2}) $ of $\su(4)^\heavy \product \su(2)_R^\heavy$.
In this case we could allow the interaction terms
\begin{align}
\cL \supset \,\,
M_{\tilde\rho}\, \overline{\tilde\rho}_L \tilde\rho_R 
+ y_{\rho\tilde\rho \Omega}\, \overline{\tilde\rho}_L \Omega_{3(1)} \rho_R + \,y_{\tilde\rho H}\, \overline{\psi}^3_{L} H^{(c)} \tilde\rho_R \, 
\end{align}
and the dimension-6 effective Yukawa couplings $Y^F_{3i}$ would be  generated, at tree-level, via the exchange of 
$\tilde\rho$ and $\rho$.

\subsection{Naturalness considerations including vector-like fermions}
\label{sec:VLF-gauge-interplay}

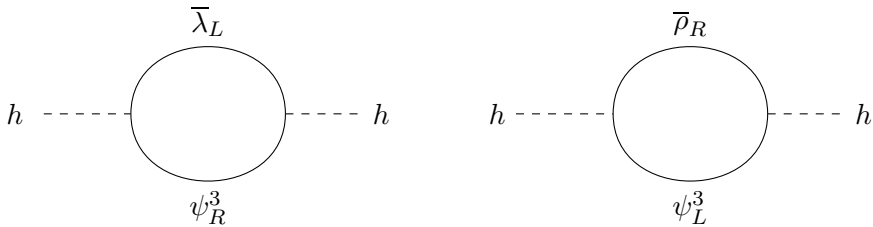
\begin{figure*}[t]
\begin{center}
\begin{tikzpicture}
\begin{feynman}
\vertex {$h$} (a);
\vertex [right=0.6in of a] (b);
\vertex [right=0.8in of b] (c);
\vertex [right=0.4in of c] (d) {$h$};
\diagram* {
(a) -- [scalar] (b) -- [plain, half left, edge label = $\overline{\lambda}_L$] (c) -- [plain, half left, edge label = $\psi_R^3$] (b),
(c) -- [scalar] (d),
};
\vertex [right=0.6in of d] (A) {$h$};
\vertex [right=0.6in of A] (B);
\vertex [right=0.8in of B] (C);
\vertex [right=0.4in of C] (D) {$h$};
\diagram* {
(A) -- [scalar] (B) -- [plain, half left, edge label = $\overline{\rho}_R$] (C) -- [plain, half left, edge label = $\psi_L^3$] (B),
(C) -- [scalar] (D),
};
\end{feynman}
\end{tikzpicture}
\end{center}
\caption{One-loop contributions to the Higgs propagator generated by the vector-like fermions $\lambda$ and $\rho$. 
\label{fig:1loopLambda} }    
\end{figure*}

As discussed in \S\,\ref{sec:VLFcompletion},  the VLFs are invariably charged under $\su(2)_L^\heavy$ and/or $\su(2)_R^\heavy$ and so couple directly to the Higgs. The VLFs therefore produce Higgs mass corrections at the one loop level, hence our naturalness criterion dictates that they cannot be too heavy.
For concreteness, consider the case of $\lambda$ and $\rho$, introduced in Eqs.~(\ref{eq:lambda}, \ref{eq:rho}) to describe flavour mixing in model 1.
The one-loop diagrams in Fig.~\ref{fig:1loopLambda} generate finite corrections to the Higgs mass scaling as 
\be \label{eq:VLF_correction}
\delta m_h^2\sim \begin{cases}
\frac{1}{16\pi^2} N_c\, |y_{\lambda H}|^2\, M_\lambda^2\,,   \\
\frac{1}{16\pi^2} N_c\, |y_{\rho H}|^2\, M_\rho^2\,, 
\end{cases}
\ee
in the limit $M_{\lambda}^2,\, M_\rho^2 \gg m_t^2,\, m_h^2$.
As expected, the result is proportional to the VLF masses squared. The naturalness condition (\ref{eq:naturalness}) then implies
\be \label{eq:naturalness_VLF}
M_\rho |y_{\rho H}|\,, \, M_\lambda |y_{\lambda H}| 
 \lesssim 800~\text{GeV}
\ee
for $\Delta \approx 1$. Taking into account the experimental bounds on vector-like fermions, which lie in the $1-2$~TeV range~\cite{CMS:2022fck},
this condition can easily be satisfied for  $|y_{\lambda H, \rho H}|  \lesssim  1/2$.

On the other hand,  the effective scales $\Lambda_{\Omega}$ and $\Lambda_R$ associated to the VLF masses must be higher than
 the symmetry breaking scales $v_\Omega$ and $v_R$ in order to generate the hierarchies in mass and mixing angles described in \S \ref{sec:mixing} 
 (and in order for the EFTs discussed in \S~\ref{sec:EFTmix} to even make sense). The one-loop naturalness conditions just derived for the VLF (\ref{eq:naturalness_VLF}) therefore translate to upper bounds on the scales $v_{\Omega,R}$, which are parametrically independent of the bounds coming from Eqs.~(\ref{eq:EWboson_correction}) and (\ref{eq:LQ_correction}). In fact we will see here that, for natural values of the VLF Yukawa couplings, these indirect naturalness constraints on $v_{\Omega,R}$ are stronger
 than those estimated previously from the direct contribution of the heavy gauge bosons to the Higgs mass.

To be concrete, we continue to discuss the particular completion via VLFs set out for model~1 in \S~\ref{sec:VLFcompletion}.
The naturalness condition (\ref{eq:naturalness_VLF}), which comes from the one-loop Higgs mass contribution from the  VLF, together with the relations $\Lambda_{\Omega} := M_\lambda/{y_{\lambda \Omega} y_{\lambda H}} = v_\Omega/\epsilon_\Omega $ and $\Lambda_{R} := M_\rho/{y_{\rho \Omega} y_{\rho H}} = v_R/\epsilon_R $, imply the following indirect naturalness constraints on $v_{\Omega}$ and $v_R$:
\begin{equation}
  v_\Omega \lesssim \frac{\epsilon_\Omega}{{y_{\lambda H}^2 |y_{\lambda \Omega}|}} \times \left( 800 \text{~GeV} \right)\, .
\end{equation}
Taking $\epsilon_\Omega \sim 0.05$ to explain the size of $|V_{cb}|$, and assuming  $|y_{\lambda H (\Omega)}| \gtrsim {0.2}$, one finds 
\be
v_\Omega = M_U/g_U \lesssim {5 \text{~TeV}\,} , 
\label{eq:MUbound2}
\ee 
where $g_U$  is the effective coupling of the heavy $SU(4)$-like gauge bosons ($g_U \approx g_4^{[3]}$ in the limit $g_4^{[3]} \gg g_3^{[12]} \approx g_s$).
Thus, by considering the origin of fermion mixing between the light and heavy generations, we are led to a naturalness `limit' on the scale of $\su(4)^\heavy \times \su(3)^\light$ symmetry breaking that is {\em lower} than the 10 TeV that was estimated using the two-loop Higgs mass correction (\ref{eq:LQ_correction}).

Similarly, but more severely, for the scale associated with the $\su(2)^\heavy$ breaking we find
\be
v_R \lesssim \frac{\epsilon_R}{\epsilon_\Omega}  \times \left( {5 \text{~TeV}} \right)\, ,
\ee
{assuming natural values $y_{\rho H (\Sigma)} \gtrsim 0.2$.}
The small parameter $\epsilon_R$ generates the hierarchy $y_2/y_3$, threrefore its natural size is $\mathcal{O}(0.01)$. 
If we take this at face value, we get a naturalness guideline $v_R = M_{W'}/\gp \lesssim {1 \text{~TeV}}$, 
which appears to be problematic from a phenomenological point of view.
What this tells us is that the degree of tuning associated with the deconstructed electroweak sector, here the $\su(2)_R$ group, 
tends to be larger than that associated with deconstructing the colour sector -- at least for this specific mechanism 
for generating the light-fermion Yukawa couplings.

We can proceed more carefully by introducing a measure 
of the tuning $\Delta$ (associated to the one-loop VLF contribution to $m_h^2$), obtaining 
\be \label{eq:refined-MW}
M_{W^\prime} \lesssim  \gp  \sqrt{\Delta} \left(\frac{\epsilon_R}{0.01}\right)  \times ({1 \text{~TeV}}) \, .
\ee
To minimize the degree of tuning needed to satisfy the phenomenological bounds on the $(W^\prime, Z^\prime)$ gauge bosons 
(see \S~\ref{sect:phenoEW} below),  one is therefore led to a region of parameter space where 
\be
 \gp  \approx g_R^\heavy \gg g_R
\ee 
is large (which in turn implies $g^\light_R \approx  g_Y^{\rm SM}$).
For concreteness, choosing $g_R^\heavy \approx {1.0}$ the naturalness bound (\ref{eq:refined-MW}) for $\Delta =1$ becomes $M_{W^\prime} \lesssim 1 \text{~TeV}$, and 
of course higher values of  $(W^\prime, Z^\prime)$ masses could be obtained tolerating a larger degree of tuning ($\Delta > 1$), {and/or by considering smaller values ($<0.2$) for the fundamental Yukawa couplings $y_{\rho H (\Sigma)}$}.
One can check that, even for this large value of $g_R^\heavy \approx 1$, the Higgs mass correction (\ref{eq:EWboson_correction}) coming from the direct one-loop gauge boson correction still satisfies $\delta m_h^2 < m_h^2$, hence the dominant source of tuning on $\delta m_h^2$ is the one associated to the VLF contributions.

We finally stress that the connection between naturalness constraints on the VLF masses and electroweak gauge boson masses
is specific to the UV completion of the EFT Lagrangian (\ref{eq:EFTmodel1}) discussed in \S \ref{sec:VLFcompletion}. 
Extending the field content, for example via heavier Higgs-like fields as in~\cite{Davighi:2022bqf},
would break this connection and leave more freedom in the natural parameter space (at the expense of having introduced more fields). 
In such cases, the $(W^\prime, Z^\prime)$ masses would be subject only to the {\em direct} naturalness constraint coming from the $\delta m_h^2$ contribution (\ref{eq:EWboson_correction}).

\section{TeV scale phenomenology}
\label{sect:phenoall}

\subsection{Strong gauge bosons}
\label{sect:pheno1}

All the interesting options in Table~\ref{tab:TeV} feature an $\su(4)^\heavy$ subgroup.
This manifestly flavour non-universal unification \`a la Pati and Salam for the third family has been proposed first in~\cite{Bordone:2017bld}, and it 
has been the subject of intensive investigations 
over the last few years~\cite{Greljo:2018tuh,Fuentes-Martin:2020pww,Fuentes-Martin:2020bnh,Fuentes-Martin:2022xnb,FernandezNavarro:2022gst,
Davighi:2022bqf,Fuentes-Martin:2019ign,Fuentes-Martin:2020luw,Fuentes-Martin:2020hvc}. The phenomenological
motivation driving these recent studies has been the intriguing {\em anomalies} in semileptonic $B$ decays (both neutral- and charged-current modes).
More precisely, these works aimed to address the $B$-physics anomalies via a class of models that also explain, in a natural way, the observed pattern of Yukawa couplings and the consistency with SM predictions of precision tests in the flavour sector.
Interestingly enough, we are led to the same unification pattern at the TeV scale 
 only by the guiding principles listed in Section~\ref{sec:assumptions},  in the absence of any phenomenological 
input from the anomalies.

The $\su(4)^\heavy$ subgroup unavoidably implies the presence of the following set of massive vectors~\cite{DiLuzio:2017vat}: the $U_1^\mu \sim (\rep{3}, \rep{1}, 2/3)$
leptoquark, the $G^\mu \sim (\rep{8}, \rep{1}, 0)$  coloron (which behaves like a heavy gluon), and a $\ZP$ gauge boson, which is a singlet under $\sm$.
Recent phenomenological studies of this suite of TeV scale gauge bosons in connection with the $B$-physics anomalies, for which the main conclusions regarding mass and coupling of the $U_1$ leptoquark are driven by the charged-current $b\to c\tau\nu$ anomaly,\footnote{In order to address the current $b\to c\tau\nu$ anomaly, $M_{U}/g_U$ should be in the range $M_{U}/g_U\in [1,2]$~TeV~\cite{Aebischer:2022oqe}.}
 can be found in Refs.~\cite{Cornella:2019hct,Cornella:2021sby,Aebischer:2022oqe,Allwicher:2023aql}. 
We cannot claim that a deviation from the SM in $b\to c\tau\nu$, at a detectable level, is an unambiguous prediction of our more general analysis in this paper:
if the $U_1$ mass saturates its naturalness bound in Eq.~(\ref{eq:MUbound2}), the phenomenological impact in $b\to c\tau\nu$ is around the $1\%$ level, and hence undetectable.
However,  our general conclusion is that this set of massive gauge bosons should appear near the TeV scale and should be mainly 
coupled to the third generation.\footnote{The so-called 4-3-2-1 non-universal gauge group, studied in detail in~\cite{Fuentes-Martin:2019ign,Fuentes-Martin:2020luw,Fuentes-Martin:2020hvc}, can well be an intermediate step, before reaching the SM, for any of the options in Table~\ref{tab:TeV}, except option 3.}
Hence an impact in $b\to c\tau\nu$ up to O(10\%) is plausible, and searches for these states at high-energies 
via $pp\to \bar\tau\tau$ ($t$--channel  $U_1$ and $s$--channel $\ZP$~\cite{Faroughy:2016osc,Baker:2019sli}) 
and $pp\to \bar t t$ ($s$--channel $G$)~\cite{Baker:2019sli,Cornella:2021sby} 
are definitely well motivated.

\subsection{Weak gauge bosons and VLF fermions} 
\label{sect:phenoEW}

In all our models we expect additional flavoured electroweak gauge bosons ($\ZP$ and $\WP$), associated to the breaking of $\su(2)_L^\heavy$ and/or $\su(2)_R^\heavy$, with masses near the TeV scale. Actually the naturalness constraints are more stringent in this case, partially because these states contribute to the Higgs mass already at the one-loop level, but also given their interplay with the mass of the vector-like fermions dictated by the flavour structure of the models (as discussed in the previous section).
As a result, the electroweak massive gauge bosons could be the lightest BSM states in our $\UU(2)^5$-inspired flavour models.
We note that the direct search constraints on these states are not particularly stringent, especially in the case of the $\su(2)_R^\heavy$ group, given the smallness 
of the couplings involved. In the absence of dedicated searches, by re-interpreting the recent CMS results in Ref.~\cite{CMS:2022ncp} we deduce a rough bound in the range $1.0$--$1.5$~TeV  for a right-handed $W^\prime$ coupled mainly to the third generation.\footnote{In our setup the $W^\prime$ production is dominated by the the $\bar b c W^\prime$ vertex, whose interaction strength is 
$g_R^\heavy \times g_Y^2/g_R^\heavy = g_Y^2$. According to the results in~\cite{Allwicher:2022gkm}, the $\bar b c$ production cross section 
is suppressed by a factor $10^{-2}$ compared to the cross section of valence quarks relevant to a SM-like $W^\prime$. 
With respect to the case analysed in~\cite{CMS:2022ncp}, the production 
cross section is therefore suppressed by a factor $10^{-2} g_Y^2/g_L^2 \approx 4\times 10^{-3}$.}

Last but not least, another generic feature of our viable models is the presence of light vector-like fermions. As shown in~(\ref{eq:naturalness_VLF}), the natural upper bounds on the masses of these 
states range from 1 TeV to several TeV, depending on the values of the corresponding Yukawa couplings.\footnote{Interestingly enough, the CMS collaboration has reported an excess in a final state 
characterised by multiple $b$ jets and $\tau$ leptons which could be interpreted as the (electroweak) 
pair-production of third-generation  vector-like leptons with mass slightly below~1 TeV~\cite{CMS:2022cpe}.}
Current bounds on the pair-production of vector-like quarks, via QCD, 
yield model-independent lower bounds on their masses of about 1.5~TeV~\cite{CMS:2022fck}. These are fully consistent 
with Eq.~(\ref{eq:naturalness_VLF}) for natural values of the Yukawa couplings.

\section{Towards the UV: general lessons}
\label{sec:UV}

We have arrived at a small number of possibilities for TeV scale new physics models with a natural electroweak sector, renormalisable Yukawa couplings for $y_t$, $y_b$, and $y_\tau$ only, and dynamics (including both new physics couplings and the light Yukawas) governed by approximate $\UU(2)$ global symmetries that follow from the $\g_U \product \g_3 \product H_{12}$ factorization of the gauge structures.  Any $\u$ factors of the gauge symmetry that talk to the Higgs and/or the top quark have already been subsumed into simple gauge factors at the low scale of several TeV.

We now discuss how these constructions could be extended further into the UV through models with flavour non-universality in the 1-2 sector,
able to reproduce the observed light flavour structure. Following the main philosophy adopted so far, we will do so considering 
models  where the electroweak scale remains natural.

\subsection{Third family compositeness} \label{sec:composite}

Before we delve into the 1-2 sector dynamics, we remark on a generic possibility for screening the electroweak scale from that UV dynamics, namely compositeness.
In all the TeV scale models that we identify, summarised in Table~\ref{tab:TeV}, the Higgs is charged as a bidoublet under one of three possible symmetries: $\su(2)_L \product \su(2)_R^\heavy$, $\su(2)_L^\heavy \product \su(2)_R$, or $\su(2)_L^\heavy \product \su(2)_R^\heavy$, all of which are isomorphic to $\mathrm{Spin}(4) \simeq \so(4)$ (where, throughout this Subsection, we use $\simeq$ to denote a local isomorphism), which we might denote by $\so(4)^\heavy$ as a reminder that this gauge symmetry is flavour non-universal. 

Recall that in most composite Higgs models, the Higgs field emerges as a pseudo Nambu Goldstone boson (pNGB) associated to spontaneous symmetry breaking of a global symmetry $\g_{\text{comp}}$ in some strongly-coupled sector, where the unbroken subgroup contains
a {\em flavour-universal}  custodial symmetry $\so(4)_U \simeq \su(2)_L \product \su(2)_R$. For example, in the minimal composite Higgs model, the global symmetry group $\g_{\text{comp}}$ is $Sp(4) \simeq \so(5)$, while other larger coset breaking patterns such as $\su(4) \simeq \so(6) \to \so(4)_U$ or $\su(4) \simeq \so(6) \to \sp(4) \simeq \so(5) \supset \so(4)_U$, have the advantage of admitting microscopic descriptions that are QCD-like. The Higgs is then naturally light, with its mass arising from explicit $\g_{\text{comp}}$-breaking effects, and being screened from dynamics above the compositeness scale. 

In the flavour non-universal models we consider here, one could embed $\so(4)^\heavy$ in the unbroken subgroup associated to a similar coset model, with a compositeness scale $\Lambda_{\text{comp}}$ not far beyond the multi-TeV scale, and where now the strong dynamics are coupled only to the third generation fermions (at least of one chirality, depending on the form of $\so(4)^\heavy$). The remaining $\g_{12}$ part of the gauge symmetry remains fundamental to higher scales. 

A proper investigation of such a third-family-compositeness approach is beyond the scope of the present work, in which we focus on the case of a fundamental Higgs up to higher scales.  We limit ourselves to remark that, if the Higgs were to arise as a pNGB in such a flavour non-universal manner, then whatever dynamics are responsible for generating the $\UU(2)$-breaking, and seeding the light family structure, could occur at energies far 
above the TeV scale, without any destabilisation of the Higgs sector.
This is an intriguing possibility phenomenologically. It would imply that the only accessible sources of flavour symmetry breaking involving 
the light families are the $\UU(2)$-breaking terms present at the TeV scale.

\subsection{The $\UU(2)$-breaking layer} 
\label{sec:U2breaking}

We now continue discussing  possible flavour non-universal  gauge dynamics that could be responsible for generating the hierarchical light 
$2\times2$ sector of the Yukawa couplings, under the hypothesis that the Higgs remains a fundamental scalar. 
Following recent explorations~\cite{Bordone:2017bld,Davighi:2022bqf}, this $\UU(2)$-breaking dynamics 
should occur at scales of order
\be
\Lambda_{12} \sim 100-1000 \mathrm{~TeV~}.
\label{eq:Lambda12} 
\ee
At this scale $H_{12}$ is embedded into a larger group, 
\be
H_{12} \hookrightarrow G_{12}\, ,
\ee
that ought to break universality between the first two generations.
Since the Higgs is not charged under $G_{12}$,  it does not couple directly to any gauge bosons in this sector.  
The Higgs only communicates to the fermions in this sector via suppressed (effective) Yukawa couplings. Thus, the scale associated with the $G_{12} \to H_{12}$ symmetry breaking can be as high as in  (\ref{eq:Lambda12})  without destabilising the Higgs mass
(the third-family-compositeness approach sketched above corresponds to $\Lambda_{12} \to \infty$, a limit which is not possible if the 
Higgs remains a fundamental field). 
Actually a scale as high as   (\ref{eq:Lambda12}) is required by flavour bounds in the 1-2 sector, coming in particular from neutral $K$-  and $D$-meson mixing.

Recall from \S\,\ref{sec:assumptions} that we eventually seek to embed our gauge group into a semi-simple one. 
While there are various possibilities for how to generate structure in the 1-2 sector, one can simplify the analysis by taking $G_{12}$ to be semi-simple (according with the notational convention introduced in footnote~\ref{footnote:semisimple}). Scales of $100-1000$ TeV are still far too low for the gauge bosons associated with $\su(5)$ or $\so(10)$ unification, hence any $\su(3)^\light$ and $\u_{B-L}^\light$ factors should be completed via $\su(4)^\light$. Taking into account this observation, and considering all 4 classes of models identified in Table~\ref{tab:TeV}, two obvious paths to the UV completion emerge.

\subsubsection{Path 1: light flavour deconstruction}

For model 1, in which $\su(2)_L$ remains flavour-universal, we can break the $\UU(2)$ symmetry by deconstructing $\u_R^\light$ by flavour. Embedding it also inside a semi-simple algebra using $\su(2)_R$ factors, we have
\be
G_{12} = \su(4)^\light \product \su(2)_R^{[1]} \product \su(2)_R^{[2]} \qquad \text{(Model 1)} .
\ee
Similarly, for model 2 in which $\su(2)_R$ remains universal, we must break $\UU(2)$ symmetries via a non-universal embedding of $\su(2)_L^\light$. The deconstruction route then gives
\be
G_{12} = \su(4)^\light \product \su(2)_L^{[1]} \product \su(2)_L^{[2]} \qquad \text{(Model 2)} .
\ee
For model 3, $\su(4)$ remains universal and the whole electroweak symmetry can be deconstructed, giving a freer structure in the 1-2 sector (with both left- and right-breaking spurions),
\be
G_{12} = \su(2)_L^{[1]} \product \su(2)_L^{[2]}\product \su(2)_R^{[1]} \product \su(2)_R^{[2]} \qquad \text{(Model 3)} .
\ee
Finally, for model 4 we have 
\be
G_{12} = \su(4)^\light \product \su(2)_L^{[1]} \product \su(2)_L^{[2]}\product \su(2)_R^{[1]} \product \su(2)_R^{[2]} \qquad \text{(Model 4)} .
\ee
In each case, a scalar `link' field charged as a bidoublet under the appropriated deconstructed $\su(2)_{L/R}$, {\em viz.}
\be
\Phi_{L/R} \sim ({\bf 2, 2}) \quad \text{of~~} \su(2)_{L/R}^{[1]} \product \su(2)_{L/R}^{[2]}\, ,
\ee
can be used to break $\su(2)_{L/R}^{[1]} \product \su(2)_{L/R}^{[2]} \to \su(2)_{L/R}^\light$ via an appropriate non-zero vev.

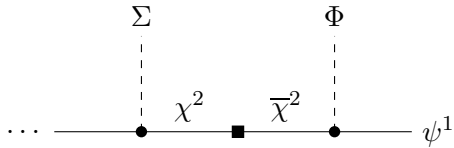
\begin{figure*}[h]
\begin{center}
\begin{tikzpicture}
\begin{feynman}
\vertex {$\dots$} (a);
\vertex [right=0.6in of a] (b);
\vertex [right=0.5in of b] (c);
\vertex [right=0.5in of c] (d);
\vertex [right=0.4in of d] (e) { $\psi^{1}$};
\node at (b) [circle,fill,inner sep=1.5pt]{};
\node at (d) [circle,fill,inner sep=1.5pt]{};
\node at (c) [square dot,fill,inner sep=1.0pt]{};
\vertex [above=0.5in of b] (f) { $\Sigma$};
\vertex [above=0.5in of d] (g) { $\Phi$};
\diagram* {
(a) -- [plain] (b)  -- [plain,edge label={ {$\chi^2$}}] (c) -- [plain,edge label={ {$\overline{\chi}^2$}}] (d) -- [plain] (e),
(b) -- [scalar] (f),
(d) -- [scalar] (g),
};
\end{feynman}
\end{tikzpicture}
\end{center}
\caption{Schematic Feynman diagram illustrating how Yukawa couplings to the first generation would be generated. Here, the $\Sigma$ field couples to a third  generation and a second generation index. The vev of $\Phi$ then links the second and first generation indices. The trailing fermion line on the left hand side has family index 3, and so can talk to the Higgs (for example, via the diagams in Fig.~\ref{fig:VLFmodel1a}). 
\label{fig:VLF_12sector} }    
\end{figure*}
 
For all the models, the $\UU(2)$-breaking in the light Yukawa sector can be achieved by supposing that the $\Sigma_L$ and/or $\Sigma_R$ fields identified in Eqs. (\ref{eq:SigmaL}--\ref{eq:SigmaR-}) come from UV fields charged as doublets under $\su(2)_{L/R}^{[2]}$ specifically, being singlets under $\su(2)_{L/R}^{[1]}$. This means that the light-light Yukawa couplings discussed in \S\,\ref{sec:EFTmix} and \S\,\ref{sec:VLFcompletion} are only generated for the second family, not the first. 
Using the scalar link fields $\Phi_{L/R}$, plus an extra VLF $\chi^2$ that has the same quantum numbers as the appropriate second generation fermion, and mass $M_{12}\sim \Lambda_{12}$ (up to some Yukawa coupling factors),\footnote{One could, instead of the VLF, use an extra vev-less scalar field with mass $M_{12}$, charged under $\su(2)_{L/R}^{[1]} \product \su(2)_{L/R}^\heavy$, which admits a cubic scalar interaction with $\Phi_{L/R}$ and $\Sigma_{L/R}$. This is more similar to the mechanism used in Refs.~\cite{Davighi:2022fer,Davighi:2022bqf}. }
one generates the Yukawa couplings to the first generation with an extra suppression factor
\be \label{eq:delta}
\delta_{L/R} \sim \frac{|\langle \Phi_{L/R} \rangle|} {\Lambda_{12}}\, 
\ee
relative to the second generation (see Fig.~\ref{fig:VLF_12sector}).

\subsubsection{Path 2: electroweak-light-flavour unification}

Interestingly, the deconstructed electroweak symmetry groups given in the previous Section are never maximal (by which we mean that they can each be embedded in a more unified, anomaly-free, semi-simple gauge theory). To wit, in every case there is also the possibility of electroweak-flavour-unification, via the embeddings:
\begin{align}
\su(2)_L^{[1]} &\product \su(2)_L^{[2]} \hookrightarrow \sp(4)_L^\light\, , \\
\su(2)_R^{[1]} &\product \su(2)_R^{[2]} \hookrightarrow \sp(4)_R^\light\, .
\end{align}
This structure also offers a natural explanation of Yukawa coupling hierarchies within the 1-2 sector, 
as discovered first in~\cite{Davighi:2022fer} (see also~\cite{Davighi:2022bqf} for a specific application to the 1-2 sector).

The identification of the $\UU(2)$-breaking spurions $\delta_{L/R}$ above can be straightforwardly lifted to the $\sp(4)_{L/R}$ models further in the UV, by embedding the link field $\Phi_{L/R}$ inside the ${\bf 5}$-dimensional representation of $\sp(4)_{L/R}^\light$, and embedding the VLF $\chi^2$ (along with a first generation partner $\chi^1$) inside the fundamental $\chi\sim {\bf 4}$ of $\sp(4)_{L/R}^\light$.

\subsection{Light Yukawa textures}

For either path, {\em i.e.}~using only flavour deconstruction or using further gauge-flavour-unification at higher energies, we generically expect the following Yukawa textures within the 1-2 sector (where we denote the $2\times 2$ Yukawa sub-matrices using the symbol $\widehat{Y}$):
\begin{align}
&\widehat{Y}^F \sim \begin{pmatrix}
\delta_R y_{11} & y_{12} \\
\delta_R y_{21} & y_{22}
\end{pmatrix} &&\text{(Model 1)}\, , \label{eq:Y12model1}\\
&\widehat{Y}^F \sim \begin{pmatrix}
\delta_L y_{11} & \delta_L  y_{12} \\
y_{21} & y_{22}
\end{pmatrix} &&\text{(Model 2)}\, , \\
&\widehat{Y}^F \sim \begin{pmatrix}
\delta_L \delta_R y_{11} & \delta_L  y_{12} \\
\delta_R y_{21} & y_{22}
\end{pmatrix} &&\text{(Models 3 \& 4)}\, ,
\end{align}
where $y_{ij}$ denote order-1 renormalisable couplings or Wilson coefficients, and $\delta_{L(R)}$ are as defined in Eq.~(\ref{eq:delta}).

The above structures can easily be put in a diagonal form via a perturbative expansion in  $\delta_{L(R)} \ll 1$. 
In model 1, 
the eigenvalues are
\be
\lambda_1 = \frac{\mathrm{Det}(y)}{y_{22}} \delta_R, \qquad \lambda_2 = y_{22}\, .
\ee
This implies that  the natural size of $\delta_R$ is $m_1/m_2$, the mass ratio between first and second generations.
The left- and right-handed  diagonalisation matrices in the singular value decomposition of 
(\ref{eq:Y12model1}) are
\be
V_L \approx \frac{1}{\sqrt{|y_{12}|^2+|y_{22}|^2}} \begin{pmatrix}
y_{22} & y_{12} \\
-y_{12}^\ast & y_{22}^\ast\, , 
\end{pmatrix}\,,
\qquad 
V_R \approx \begin{pmatrix} 1 & \delta_R \\ -\delta_R & 1 \end{pmatrix}\, .
\ee
From these expression we deduce that in this model the Cabibbo angle is not parametrically suppressed.
This works well phenomenologically:  it is entirely natural to generate the observed value of the 
Cabibbo angle ($\theta_c \approx 13^{\circ}$) from $y_{ij}$ entries of $O(1)$ in both up- and down-type Yukawa
couplings. On the other hand, in this model the mixing in the right-handed sector is parametrically suppressed by  $\delta_R \sim m_1/m_2 \sim 10^{-2}$. This  is also quite welcome, given the strong experimental constraints on flavour-violating right-handed currents in the light quark sector. 

The situation for models of type 2 is the reverse. The mass hierarchies would now 
imply $\delta_L \sim m_1/m_2$, with the Cabibbo angle correspondingly suppressed by the same factor
and no parametric suppression present in right-handed mixing.  
Similarly to what was already observed in the light-heavy mixing (\S\,\ref{sec:mixing}), this parametric structure can only be made compatible 
with observations with an unnatural fine-tuning of the $O(1)$ parameters. 

Models of type 3 and 4 are, again, occupying the middle ground. For these, the mass ratios are now doubly-suppressed via
\be
\frac{m_1}{m_2} \sim \delta_L \delta_R\, ,
\ee
while $\delta_L$ and $\delta_R$ separately determine the mixing angles appearing in $V_L$ and $V_R$, respectively. Thus, the parameter $\delta_R$ can be set to the mass ratio, while $\delta_L$ is set to the Cabibbo angle. In other words, accounting for the data requires a small hierarchy of scales,
$\delta_R \ll \delta_L$, in models of types 3 and 4, echoing Eq. (\ref{eq:eps_hierarchy}), which is entirely reasonable.

\section{Flavour-universal $\su(2)_L$ in the UV: a complete model}
\label{sec:UVwinning}

From the discussions of the previous Sections one class of models emerges as the most economical and natural in satisfying 
all the hypotheses listed in Section~\ref{sec:assumptions}, namely the  model 1 in Table~\ref{tab:TeV}, where 
$\su(2)_L$ remains flavour universal. 
The model 4 is a valid, slightly less economical, alternative; however, in this case specific UV completions have already been discussed in the literature~\cite{Bordone:2017bld,Fuentes-Martin:2022xnb,Davighi:2022bqf}. 
For that reason, we here focus on a specific UV completion of model~1, continuing to employ the (strong) assumption that  the Higgs 
remains a fundamental field up to the scale where the full gauge symmetry is embedded  into a semi-simple group.

Following the general arguments presented in preceding Sections, we construct a UV theory with the semi-simple gauge group 
\be
\g = \su(2)_L \product \su(4)^\heavy \product \su(4)^\light \product \su(2)_R^\heavy \product \sp(4)_R^\light\, .
\ee
We denote the gauge couplings for these five factors by 
\be \label{eq:couplings}
\{g_L, g_4^h, g_4^l, g_R^h, g_R^l\}\,,
\ee
where the labels $h$ and $l$ here stand for `heavy' and `light'.
Taking into account the results of \S\,\ref{sec:mixing} and \S\,\ref{sec:U2breaking}, 
the Yukawa coupling matrices have the following parametric dependence
\be
Y^F \sim \begin{pmatrix}
\epsilon_R \delta_R & \epsilon_R & \epsilon_\Omega \\
\epsilon_R \delta_R & \epsilon_R & \epsilon_\Omega \\
\epsilon_R \epsilon_\Omega \delta_R & \epsilon_R \epsilon_\Omega & 1
\end{pmatrix}\, ,
\ee
where the small parameters are defined as
\be
\epsilon_\Omega = \frac{\langle\Omega\rangle}{\Lambda_\Omega}, \qquad 
\epsilon_R = \frac{\langle \Sigma_R \rangle}{\Lambda_R}, \qquad
\delta_R = \frac{\langle \Phi_R \rangle}{\Lambda_{12}}\,.
\ee
This implies 
\be
\frac{y_1}{y_2} \sim \delta_R, \qquad \frac{y_2}{y_3} \sim \epsilon_R, \qquad |V_{cb}| \sim \epsilon_\Omega, \qquad |V_{us}| \sim \mathcal{O}(1)\, .
\ee
The same parametric scaling holds for the mass ratios and the mixing angles in the charged lepton sector.

\subsection{Splitting the bottom and tau Yukawa couplings} \label{sec:QLsplitting}

While these parametric scalings of fermion masses and mixing angles offer a good (and economical) description of the overall hierarchical pattern, it is important to check that the numerical differences between charged lepton and quark masses can be accommodated in the model. This is not automatic due to the quark-lepton unification. In particular, the $\su(4)^\heavy$ symmetry (which is present not only in model 1,  but also in models 2 and 4), 
predicts equality between bottom and tau Yukawas at the TeV scale (and above). This is nearly, but not quite achieved by 
the renormalization group evolution of these coupling within the SM, which yields $m_\tau  (\mu) = 1.78$ GeV and $m_b(\mu) = 2.43 \pm 0.08$ GeV at $\mu = 1$ TeV~\cite{Xing:2007fb}.

This is a well-known phenomenological problem in BSM models using low-scale Pati--Salam unification (either in the third family, or in all families).
As we discuss below, there are several possible solutions, each of which requires adding extra matter fields.

\paragraph{Option 1: {\boldmath$H_{15}$}.}
Following the more traditional path, the Pati--Salam matter content can be extended by adding an extra Higgs field in the representation $H_{15}\sim ({\bf 15, 2, 2})$ of $\su(4)^\heavy \product \su(2)_L \product \su(2)_R^\heavy$. This allows one to build a second set of Yukawa couplings, of the form $\overline{\psi}_L^3 H_{15} \psi_R^3$, that 
can distinguish quarks from leptons via a vev of $H_{15}$ aligned to the $\su(4)$ adjoint generator $T_{15} \propto \text{diag}(1,1,1,-3)$.

\paragraph{Option 2: {\boldmath$\Omega_{15}$} and fermion mixing.}
In the presence of the VLF $\lambda$ (\ref{eq:lambda}), which has the same quantum numbers as the chiral (would-be-SM) field $\psi_L^3$, there can be mixing between $\psi_R^3$ and the right-handed component of $\lambda$. 
Including also an extra $\R$-valued scalar field $\Omega_{15}$, charged in the adjoint representation of $\su(4)^\heavy$ and neutral under everything else, the  gauge-eigenbasis Lagrangian admits the terms~\cite{DiLuzio:2018zxy,Cornella:2019hct}
\be \label{eq:Om15sector}
\cL \supset \overline{\lambda}_L\left( M_\lambda + y_{15} \Omega_{15} \right) \lambda_R + \tilde{m} \overline{\lambda}_L \psi_R^3 \, .
\ee
Assuming that $\Omega_{15}$ acquires a non-zero vev 
\be 
\langle \Omega_{15} \rangle = v_{15} \,\, \text{diag}(1,1,1,-3)
\ee
at the TeV scale, thereby contributing to $\su(4)^\heavy$-breaking, the coupling $y_{15}$ gives rise to a mass-splitting  between the quark and lepton components of the $\su(4)^\heavy$ ${\bf 4}$-plets in $\lambda$. Because of the mixing between this VLF and the third family RH chiral field (via the coupling $\tilde m$), this induces a splitting between the third family Yukawa couplings of quarks and leptons.

To see this quark-lepton mass-splitting effect, let us decompose the fermion fields into their quark and lepton components. Working with $\su(2)_R^\heavy$ symmetry still manifest, we write $\psi_R^3 = (q_R^3, l_R^3)$ for the RH chiral field. Let $\lambda_{L/R} = (Q_{L/R}, L_{L/R})$, where $Q$ are the quark components ($\su(3)^\heavy$ triplets) and $L$ are the leptonic ones.
The RH fields that remain massless above the scale of electroweak symmetry breaking are then
\begin{align}
q_R^{3\prime} &:= -\sin\theta_q \,Q_R + \cos \theta_q\, q_R^3\, , \\
l_R^{3\prime} &:= -\sin\theta_l \,L_R + \cos \theta_l\, l_R^3\, , 
\end{align}
where
\be
\sin\theta_{q} = \frac{\tilde m}{\sqrt{(M_\lambda + y_{15}v_{15})^2+\tilde{m}^2}}\,, \qquad
\sin\theta_{l} = \frac{\tilde m}{\sqrt{(M_\lambda -3 y_{15}v_{15})^2+\tilde{m}^2}}\,.
\ee
Hence the Yukawa couplings of the massless chiral fermion components to the Higgs take the form
\be
\cL \supset y_3 \, \overline{\psi}_L^3 H \psi_R^3 \supset y_3 \left( c_q \overline{q}_L^3 H q_R^{3\prime} + c_l \overline{l}_L^3 H l_R^{3\prime} \right)\, ,
\ee
where we used the shorthand notation $c_{q,l} :=\cos\theta_{q,l}$ and $s_{q,l} :=\sin\theta_{q,l}$. 
The freedom to achieve $c_q \neq c_l$ via $y_{15} \neq 0$ then allows us, after the breaking of $\su(4)^\heavy$, to obtain the desired splitting $m_\tau \neq m_b$.  

\paragraph{Option 3: additional VLF.}

In option 2, we assume that the field $\Omega_{15}$ is a physical, dynamical field, hence the Yukawa interaction in Eq.~(\ref{eq:Om15sector}) is a dimension-four term. As an alternative, the same interaction can arise from an effective dimension-five coupling that involves the other fields already in the model, namely
\be \label{eq:OmOm_operator}
\cL_5 \supset \frac{1}{M} \overline{\lambda}_L (\Omega_4 \Omega_4^\dagger) \lambda_R\, , 
\ee
where $M$ is some heavy scale, and $\Omega_4= (\Omega_3, \Omega_1)$ unifies $\Omega_3$ and $\Omega_1$, as defined in (\ref{eq:Om3}), after the embedding $\su(3)^\light \product \u_{B-L}^\light \hookrightarrow \su(4)^\light$ at the $\Lambda_{12}$ scale. By construction, 
$\Omega_4$ transforms in the representation $({\bf \overline{4}, 4})$ of $\su(4)^\heavy \product \su(4)^\light$. A simple UV completion of 
the effective interaction in Eq.~(\ref{eq:OmOm_operator}) is to include one more VLF, in the representation 
\be
\tilde \lambda \sim ({\bf 2, 4}) \quad \text{of~~} \su(2)_L \product \su(4)^\light\, ,
\ee
and an interaction $\cL \supset \overline{\lambda}_L \Omega_4 \tilde{\lambda}_R$, which generates the operator (\ref{eq:OmOm_operator}) 
via the tree-level exchange of $\tilde{\lambda}$.

\begin{table}[t]
\begin{center}{\small
\begin{tabular}{|c|c|c|c|cc|cc|}
\hline
 & Field & $J$ & $\su(2)_{L}$ & $\su(4)^\heavy$ & $\su(4)^\light$ & $\su(2)_R^\heavy$ & $\sp(4)_R^\light$ \\
\hline
EWSB Higgs & $H_1$ & $0$ & $\bf{2}$ &  &  & $\bf{2}$ &  \\
\hline
\rowcolor{LightGray}
$m_\tau$ {\em vs.} $m_b$ splitting & $\Omega_{15}^{\R}$ & $0$  & & $\bf{15}$ & & & \\
\hline
Chiral (SM-like)  & $\psi_L^3$ & $1/2$  & $\bf{2}$ & $\bf{4}$ & & & \\
Fermions	& $\psi_R^3$ & $1/2$& & $\bf{4}$ & & $\bf{2}$ &   \\
	& $\psi_L^{i\in\{1,2\}}$ & $1/2$ & $\bf{2}$ &  & $\bf{4}$ & &  \\
	& $\psi_R^l$ & $1/2$ &  && $\bf{4}$ &  & $\bf{4}$  \\
\hline
\rowcolor{LightGray}
TeV sector & $\Omega_4$ & $0$  & &  $\overline{\bf{4}}$ &  $\bf{4}$ & & \\
\rowcolor{LightGray}
	& $\Sigma_R$ & $0$  & &&&  $\bf{2}$ &  $\bf{4}$ \\
\rowcolor{LightGray}
	& $\lambda_{L/R}$ & $1/2$  &  $\bf{2}$ & ${\bf{4}}$ & & & \\
\rowcolor{LightGray}
	& $\rho_{L/R}$ & $1/2$ & & & ${\bf{4}}$ &$\bf{2}$ & \\ 
\hline
\rowcolor{LightGray}
$\UU(2)$-breaking & $S_R$ & $0$  &   & & $\overline{\bf{4}}$ &  & $\bf{4}$  \\
\rowcolor{LightGray}
sector & $\Phi_R$ & $0$  & & & & &  $\bf{5}$ \\
\rowcolor{LightGray}
	& $\chi_{L/R}$ & $1/2$ & & & $\bf{4}$  & &  $\bf{4}$ \\
\hline
\end{tabular}
}
\end{center}
\caption{Field content for a possible UV completion of model 1. The shaded rows indicate BSM field content, and the column labelled $J$ records the spin (scalar or Weyl fermion) of each matter field. } \label{tab:UVmodel}
\end{table}

Of these three, options 2 and 3 appear more interesting: they involve fewer fields and, importantly, no proliferation of SM Higgs-like fields is needed. 

\subsection{Complete model}

For the sake of writing down a definite UV completion, we choose to proceed with option~2. One good reason for this 
is that only in this case are the one-loop $\beta$-functions for all five gauge couplings in the model negative, meaning that the gauge interactions flow towards zero in the ultraviolet (see \S\,\ref{sec:AF} below).

The resulting field content of the model, now UV complete, is shown in Table~\ref{tab:UVmodel}. Various other choices have been made, especially in the 1-2 sector, guided by minimality.
For example, the scalar field $S_R \sim (\repbar{4}, \rep{4})$ of $\su(4)^\light\product \sp(4)_R^\light$ condenses at a high scale to trigger a symmetry breaking
\be \label{eq:12-high-breaking}
\su(4)^\light\product \sp(4)_R^\light \longrightarrow \su(3)^\light \product \su(2)_{R}^{[1]} \product \u_R^\prime\, ,
\ee
where $\u_R^\prime$ acts as $B-L$ on the first family fermions, and hypercharge on the second family,
as in Ref.~\cite{Davighi:2022bqf} (see also~\cite{Davighi:2022fer}). Under (\ref{eq:12-high-breaking}), the complex scalar field $\Phi_R \sim {\bf 5}$ of $\sp(4)_R^\light$ decomposes into two doublets of $\su(2)_R^{[1]}$ with opposite $\u_R^\prime$ charges (plus a singlet). These doublets then condense to link together $\su(2)_R^{[1]} \product \u_R^\prime \to \u_Y^\light$, resulting in the light-family colour and hypercharge symmetries of the SM (recall that $\su(2)_L$ remains family universal at all scales in this model). The VLF $\chi$, in particular its second generation component, then mediates the Yukawa couplings to first generation right-handed fields in the manner described in \S \ref{sec:U2breaking}.

\subsection{Asymptotic freedom of the gauge couplings} \label{sec:AF}

As anticipated, an interesting property of the UV complete model we have identified is the asymptotic freedom of all its gauge couplings. 
Using the textbook formula for the one-loop $\beta$-functions 
\be
\beta_i := \frac{\partial g_i}{\partial \ln \mu} = -\frac{g_i^3}{16\pi^2}\left[ \frac{11}{3} C_2({\bf Ad}_{\g_i}) - \frac{2}{3} \sum_{\{\text{Weyls},~ F\}} T({\bf r}_F) - \frac{1}{6} \sum_{\{\R \text{~scalars},~S\}} T({\bf r}_S) \right]\, ,
\ee
where ${\bf Ad}_{\g_i}$ denotes the adjoint representation of the factor $\g_i$ (for which $g_i$ is the associated gauge coupling), $C_2({\bf R})$ denotes the quadratic Casimir of a representation ${\bf R}$ and $T({\bf R})$ denotes its Dynkin index,
we obtain the results shown in Table~\ref{tab:beta}. For completeness, we report the results for each of the three options discussed in \S\,\ref{sec:QLsplitting}.
As can be seen, only for our chosen option~2 are all the $\beta$-functions negative. 

Having identified a UV completion where all the gauge $\beta$-functions are negative is quite remarkable:
such a model has a chance of flowing to a weakly coupled theory in the UV, at least in the gauge sector, adding to its appeal as a fundamental theory. We stress that this feature is relatively uncommon amongst fully-fledged models of flavour. 
The fact that $\su(2)_L$ is left flavour-universal up to the UV, with very little BSM matter coupled to it (only the VLF $\lambda$), is an important feature that renders this asymptotic behaviour possible. 

\begin{table}
\begin{center}
\begin{tabular}{|c|c|c|c|}
\hline
$g_i$ & $\frac{16\pi^2}{g_i^3}\beta_i$ (with $H_{15}$) & $\frac{16\pi^2}{g_i^3}\beta_{i}$ (with $\Omega_{15}^{\R}$)  & $\frac{16\pi^2}{g_i^3}\beta_i$ (with $\tilde\lambda$)  \\
\hline
$g_L$ &  $+14/3$ & $-1/3$ & $+7/3$\\
$g_4^h$ & $-6$ & $-32/3$ & $-34/3$\\
$g_4^l$ & ${-20/3}$ & ${-20/3}$ & ${-16/3}$\\
$g_R^h$ & $+8/3$ & $-7/3$ & $-7/3$\\
$g_R^l$ & ${-17/3}$ & ${-17/3}$ & ${-17/3}$\\
\hline
\end{tabular}
\end{center}
\caption{One-loop $\beta$-functions, for the gauge couplings, in the UV completions of model 1. The different columns correspond to different 
options employed to address the bottom-tau mass splitting (see text). \label{tab:beta}}
\end{table}

\subsection{Partial Higgs seclusion}

To conclude our discussion about this UV complete model, it is worth revisiting the stability of the Higgs sector. 
By keeping $\su(2)_L$ flavour-universal up to the UV, the model benefits from an in-built `protection' of the Higgs mass, firstly by avoiding heavy gauge bosons  charged under $\su(2)_L$, and secondly by avoiding the need for further $\su(2)_L$-charged states in order 
to generate fermion mixings. Interestingly, this is the same feature that allowed for the model to have a negative beta function for the $\su(2)_L$ gauge coupling. 

The only $\su(2)_L$-charged BSM state in the theory is the VLF $\lambda$, that we expect to satisfy the naturalness condition 
(\ref{eq:naturalness_VLF}).
Assuming the Yukawa coupling $y_{\lambda H}$ to be some modest fraction of $y_t$, say $y_{\lambda H} \sim 1/3$, we see that a VLF mass around 2 TeV (well consistent with present data) would be perfectly natural. A similar conclusion holds for the second VLF $\rho$. 

The other exotic states that couple directly to the Higgs field are the heavy electroweak gauge bosons, also expected to be close to 1~TeV in the absence of fine tuning
(see \S~\ref{sec:VLF-gauge-interplay}). For this particular UV complete model, these electroweak states originate only from the deconstruction of $\su(2)_R$. 
There is an added double benefit in deconstructing $\su(2)_R$, rather than $\su(2)_L$, which is provided by the smallness of $g_R$ with respect to $g_L$. This fact implies firstly that, at fixed symmetry-breaking scale, the induced Higgs-mass corrections are accordingly smaller (or, equivalently, the  symmetry-breaking scale can be higher at fixed Higgs-mass correction). Secondly, the smallness of  $g_R$, and of the right-handed mixing angles in the fermion sector, implies weaker bounds from 
direct searches (via couplings to light fermions), as discussed in Section~\ref{sect:phenoall}.

No other exotic states affect the Higgs mass at the one-loop level. The corrections due to the $\su(4)^\heavy$ gauge bosons, 
which appear at the two-loop level, imply upper bounds on the corresponding heavy gauge boson mass of a few TeV
(as discussed in \S~\ref{sect:NatG} and \S~\ref{sec:VLF-gauge-interplay}).
Finally, one might worry about the effects of the $\su(4)^\light$ gauge bosons, whose masses must be at the higher scale $\Lambda_{12} \sim 1000$ TeV due to flavour bounds. However, these gauge bosons couple to the Higgs only via the mediation 
of second generation SM fermions, so their impact on the Higgs mass features a light-Yukawa suppression ($y_2/y_3$) 
which compensates for the ratio of scales $\Lambda_{12}/\Lambda_3$. Similar considerations apply to the $\sp(4)_R^\light$  gauge bosons.
We can thus conclude that in this model the Higgs field is quite effectively `secluded' with respect to the heavy modes.

\section{Summary and conclusions}

Flavour is a rich and puzzling source of structure in particle physics, that is unexplained in the Standard Model (SM). Dynamical solutions to the flavour puzzle generically entail heavy particles with large couplings to the third family, which therefore talk strongly to the Higgs because of the large top-quark Yukawa coupling. This means that theories of flavour, unless carefully crafted, are likely to induce large quantum corrections to the Higgs mass and so destabilize the electroweak scale. Interpreting this differently, we would argue that there is an inescapable link between Higgs and flavour that we should not try to disentangle, but rather take seriously in our exploration of physics beyond the SM.

In this work, we systematically explored a class of new physics models that offer a complete and elegant explanation of the SM flavour structure, via flavour non-universal gauge interactions, but which entail only small quantum corrections to the Higgs mass. We focused on models described by weakly interacting non-supersymmetric gauge theories, with the Higgs remaining a fundamental field well above the TeV scale. We also restricted to theories that embed inside a semi-simple gauge group (without extra chiral fermions), thereby also seeking an explanation of hypercharge quantization, and even offering a shot at asymptotically free dynamics in the UV. 

We began by using naturalness to estimate upper bounds on the scale associated with various flavour-non-universal unification patterns, and thence ruling out options which are clearly incompatible with data at their `natural scale'. This excludes $\su(5)$-based unification patterns due to the proton decay bound, and also disfavours models with a flavour-universal quark-lepton $\su(4)$ symmetry, which must be broken above 200 TeV to not contradict constraints on {\em e.g.} $\mu \to e$ violation. These combined conditions single out a surprisingly small class of viable 
TeV-scale models, summarised in Table~\ref{tab:TeV}. Their most striking common characteristic is quark-lepton unification in the third family via an $\su(4)^\heavy$ symmetry, which implies the existence of a $U_1$ vector leptoquark with mass in the few TeV domain.

We explored in more detail the flavour structure of all the allowed options, where we required also part of the electroweak symmetry to be  deconstructed in the multi-TeV domain in order to reproduce the approximate $\UU(2)^5$ flavour symmetries observed in the 
Yukawa couplings. Restricting ourselves to minimal sets of symmetry-breaking scalar fields and vector-like fermions as extra ingredients, we extended each model deeper into the UV in order to fully resolve the light-fermion sector. This analysis shows that, besides the class of models studied in~\cite{Bordone:2017bld,Fuentes-Martin:2022xnb,Davighi:2022bqf}, an interesting and more `economical' alternative 
is the possibility of keeping $\su(2)_L$ flavour-universal up to high scales, with $\su(2)_R$ deconstructed to seed the Yukawa hierarchies. This natural and elegant possibility, which has escaped attention in the literature until now, predicts suppressed mixings in the right-handed sector (both between light and heavy generations, and also within the light generations) without any need for a hierarchy of associated symmetry breaking scales. In other words, this gauge structure not only gives rise to approximate accidental  $\UU(2)^n$ flavour symmetries, it also 
efficiently realises its minimal breaking, as defined in~\cite{Barbieri:2011ci,Isidori:2012ts}, without any tuning.

We concluded our study by postulating a UV complete model within this class. By fully explaining the SM flavour structure while simultaneously keeping the electroweak scale stable, the model necessarily predicts new particles near the TeV scale that will be probed by current and future experiments: a vector-like fermion with mass in the 1-2 TeV range, a set of (flavour non-universal) right-handed $W^\prime$ and $Z^\prime$ bosons with mass in the same range and, as for all other models, a third-family-aligned vector leptoquark with mass in the few TeV domain. Of course, higher masses are viable, but at the expense of increased fine-tuning.
As a bonus, we even found that the model has negative one-loop beta functions for all its gauge couplings, which is a rare feature in such complete models of flavour. 

The details of the UV completions we have proposed are certainly highly speculative. For instance, other stabilisation mechanisms of the Higgs  sector such as supersymmetry or compositeness could step in at intermediate energy scales, changing the UV behaviour. 
However, our general conclusions about TeV-scale dynamics sketched in Section~\ref{sect:phenoall}  
are based on the more robust hypothesis of gauge non-universality for the third family, and reconciling this hypothesis with Higgs mass stability. 
This represents a shift of paradigm compared to the old Minimal Flavour Violation hypothesis~\cite{DAmbrosio:2002vsn}, which was based on the implicit assumption of gauge universality. Our main message is that this new paradigm emerges, in view of present high-energy and flavour data, as a well-motivated option for addressing the inescapable link between Higgs and flavour.

\acknowledgments 

We are grateful to R. Barbieri for providing valuable feedback on the first version of this manuscript.
This work is funded by the European Research Council (ERC) under the European Union’s Horizon 2020 research and innovation programme under grant agreement 833280 (FLAY), and by the Swiss National Science Foundation (SNF) under contract 200020-204428.

\bibliographystyle{JHEP}
\bibliography{refs}
\end{document}